 \documentclass[usenatbib]{mn2e}
\usepackage{graphicx}
 \usepackage{epstopdf}
 \usepackage{amsmath,amssymb}

\newcommand{\apj}{ApJ}
\newcommand{\apjl}{ApJ}
\newcommand{\apjs}{ApJS}
\newcommand{\aap}{A \& A}
\newcommand{\aj}{AJ}
\newcommand{\mnras}{MNRAS}
\newcommand{\physrep}{Physics Reports}
\newcommand{\prd}{Phys Rev D}
\newcommand{\nat}{Nature}

\newcommand{\lya}{Ly$\alpha$ }
\newcommand{\lyf}{Ly$\alpha$ forest }
\newcommand{\lyfns}{Ly$\alpha$ forest}
  \newcommand{\iMpc}{\mbox{ Mpc$^{-1}$}}
   
      \newcommand{\tauCMB}{$\tau_{\rm CMB}$ }   
               
            \newcommand{\gammaI}{$\Gamma_{-12}$ }  
                        \newcommand{\gammaIns}{$\Gamma_{-12}$}  
                        \newcommand{\tauEff}{$\tau_{\rm eff}$ } 
                           
      \newcommand{\Ndot}{$\dot{N}_{\rm ion}$ }  
            \newcommand{\Ndotns}{$\dot{N}_{\rm ion}$}  
   \newcommand{\K}{\mbox{ K}}

   \newcommand{\ud}{\mbox{d}}   
   
   \defcitealias{me2000}{MHR00}
   \defcitealias{bolton2007}{BH07}
   \defcitealias{fg2008}{FG08}   

      \title[Observational constraints on reionization and 21cm experiments]{Constraining reionization using 21 cm observations in combination with CMB and Lyman-alpha forest data}
 
 \author[Jonathan R. Pritchard, Avi Loeb, and Stuart Wyithe]{Jonathan R.~Pritchard$^1$
\thanks{Hubble Fellow; Email: jpritchard@cfa.harvard.edu}, Abraham Loeb$^1$, and J.~Stuart~B.~Wyithe$^2$\\ 
 $^1$Harvard-Smithsonian Center for Astrophysics, MS-51, 60 Garden St, Cambridge, MA 02138 \\
 $^2$School of Physics, University of Melbourne, Parkville, Victoria, Australia}

\begin{document}
 
\maketitle

 \begin{abstract}
In this paper, we explore the constraints on the reionization history that are provided by current observations of the \lya forest and the CMB. Rather than using a particular semi-analytic model, we take the novel approach of parametrizing the ionizing sources with arbitrary functions, and perform likelihood analyses to constrain possible reionization histories. We find model independent conclusions that reionization is likely to be mostly complete by $z=8$ and that the IGM was 50\% ionized at $z=9-10$. Upcoming low-frequency observations of the redshifted 21 cm line of neutral hydrogen are expected to place significantly better constraints on the hydrogen neutral fraction at $6\lesssim z\lesssim12$. We use our constraints on the reionization history to predict the likely amplitude of the 21 cm power spectrum and show that observations with the highest signal-to-noise ratio will most likely be made at frequencies corresponding to $z=9-10$. This result provides an important guide to the upcoming 21 cm observations. Finally, we assess the impact that measurement of the neutral fraction will have on our knowledge of reionization and the early source population. Our results show that a single measurement of the neutral fraction mid-way through the reionization era will significantly enhance our knowledge of the entire reionization history.

\end{abstract}

 

\section{Introduction} 
\label{sec:intro}

Determining how and when the Universe was reionized is one of the major outstanding questions of modern cosmology.  Observations of the cosmic microwave background confirm that hydrogen gas first became neutral at $z\approx1100$ around 400,000 years after the Big Bang.  Observations of absorption spectra of quasars at redshifts $z<6$ indicate that the intergalactic medium (IGM) is largely ionized.  Therefore a cosmic phase transition occurred whereby the IGM went from being fully neutral to being almost fully ionized \citep[for a review see][]{barkana2001}.  Our picture of when exactly this phase transition occurred and what was the nature of the sources that drove it remains cloudy.  

In recent years, improved observational data has begun to probe the reionization era.  By pursuing observations of the \lya forest in QSO spectra at ever higher redshifts, Gunn-Peterson troughs \citep{gp1965,scheuer1965} have been seen along several quasar sight lines at $z\gtrsim6$ \citep{becker2001}, providing an indication of an increase in the IGM neutral fraction at these redshifts.  Interpretation of these observations is subtle, since the IGM ionization state depends upon its thermal history, the spectrum of the ionizing sources, and the history of the production of ionizing radiation in those sources.

Complementing \lya forest observations, is the detection of the imprint of reionization in the polarization of the cosmic microwave background (CMB).  Rescattering of CMB photons by the ionized IGM produces a distinctive ``bump" in the CMB power spectrum on large angular scales that has been detected by WMAP \citep{dunkley2009}.  This places a constraint on the integrated optical depth \tauCMB to the surface of last scattering, and suggests that a significant fraction of the IGM was ionized by $z\approx10$.   In a few years time, results from the Planck satellite\footnote{http://www.rssd.esa.int/index.php?project=planck} are expected to tighten these constraints considerably.

In addition to these relatively robust observations, a host of other observations contribute to our understanding of the state of the IGM at high redshift.  These include constraints on the temperature of the IGM in the range $z=2-4$ \citep{schaye2000,ricotti2000,mcdonald2001,zaldarriaga2001,cen2009}.  These observations are challenging and their interpretation is complicated by the likelihood that HeII reionization is occurring at $z\approx3$.
Other observations include the evolution of the abundance and clustering of \lya emitters \citep{malhotra2004} and of the evolution of the metallicity \citep{becker2009}.  These observations are difficult to interpret and as of yet offer only hints rather than definite evidence for how reionization proceeds.  There also exist constraints on the number density of galaxies at high redshift \citep{bouwens2005,stark2007}, although extrapolating from the few galaxies observed to their implications on the ionization history is difficult due to the small number of candidate high redshift galaxies and the unknown escape fraction of ionizing photons.

In the near future, it is hoped that observations of the redshifted 21 cm line of neutral hydrogen will provide a direct probe of the state of the IGM at redshifts $z=6-15$.  Low frequency interferometers such as LOFAR\footnote{http://www.lofar.org/}, MWA\footnote{http://www.MWAtelescope.org/}, PAPER\footnote{\citet{parsons2009}}, and SKA\footnote{http://www.skatelescope.org/} should, in principle, be able to map the ionization state of the IGM giving direct measurements of the ionized fraction $x_i$ at several different redshifts.  It is also hoped that future observations of high-redshift gamma ray bursts (GRB) will illuminate the state of the \lya forest at yet higher redshifts \citep{bromm2006,mcquinn2008grb} and the recent observation of a $z\approx8$ GRB \citep{tanvir2009} seems cause for optimism.

On the theoretical side, numerical simulations offer one way of trying to reconcile these different observations and a variety of groups have made concrete progress in this area \citep[for a recent review see ][]{trac2009}.  Analytic modelling is also possible and provides a useful way of exploring a large parameter space in order to assess how different sources of information constrain reionization.  For example, \citet{choudhury2005,choudhury2006} attempt to simultaneously model many of the above observations with a single model in a self-consistent way.  Despite progress, there is considerable uncertainty in the theoretical modelling of reionization, partly stemming from uncertainty in the relevant parameters and partly from the complexity of the interaction between different physical processes.

In this paper, we combine these different pieces of observational evidence in order to quantify our uncertainty on the reionization history.  Given the numerous sources of astrophysical uncertainty, we argue that it is presumptive to claim that we can use analytic models to definitively constrain reionization in detail (e.g. Choudhury \& Ferrara~2005).  Instead we propose arbitrary forms for the evolution of ionizing photon production.  Using these arbitrary forms, we define the space of plausible models that fit the current data.  We focus on the \lyf and CMB observations which most tightly constrain reionization, and which are most readily predicted by an analytic model. We use a likelihood based analysis to place constraints on the ionization history, considering two different source modelling parametrizations, in order to assess the systematic uncertainty in our conclusions.

Having made explicit the bounds of our ignorance in the ionization history, we explore the implications for the signal-to-noise ratio (S/N) that future 21 cm experiments might achieve at different redshifts.  Finally, we turn this problem around and ask what 21 cm experiments will tell us about the reionization history that was not already implicit in our existing observations.

The layout of this paper is as follows.  In \S\ref{sec:observations} we discuss the current observational constraints on reionization that we will be using in our analysis.  We then discuss in \S\ref{sec:inference} the formalism for using these constraints to infer the ionization history given a particular model for reionization.   In \S\ref{sec:results}, we use this formalism to place constraints on model parameters and calculate probability distributions for the neutral fraction in different redshift bins.  This is then used to make predictions for upcoming 21 cm instruments.  Finally, we conclude in \S\ref{sec:conclude}.  The explicit details of our model of reionization are left to appendix \ref{sec:modeling}.

Throughout this paper, we assume a cosmology with $\Omega_m=0.3$, $\Omega_\Lambda=0.7$, $\Omega_b=0.046$, $H=100h\,\rm{km\,s^{-1}\,Mpc^{-1}}$ (with $h=0.7$), $n_S=0.95$, and $\sigma_8=0.8$, consistent with the latest measurements \citep{komatsu2009}.

\section{Observational constraints on reionization} 
\label{sec:observations}

We now turn to the existing observational constraints on reionization. Constraints on reionization fall into two main types: those, like $\tau_{\rm CMB}$, that depend solely upon the ionization history $x_i(z)$ and those, like the \lyfns, that constrain the sources and ionizing background that drives reionization. In this section, we discuss existing and future constraints and specify the data sets that we will use for our inference.  We defer the details of our model of reionization to Appendix \ref{sec:modeling}.

\begin{figure}
\begin{center}
\includegraphics[scale=0.4]{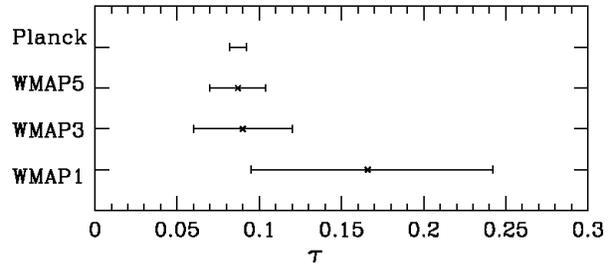}
\caption{Evolution of WMAP $1-\sigma$ constraints on the optical depth $\tau_{\rm CMB}$ and, for comparison, the predicted error for Planck.}
\label{fig:tauCMB}
\end{center}
\end{figure}
Perhaps the most robust constraint on reionization comes from WMAP measurements of the optical depth to the surface of last scattering.  Figure \ref{fig:tauCMB} graphically illustrates the evolution of measurements of \tauCMB.  Following the inclusion of improved polarisation measurements in WMAP3, the best fit value appears to have stabilised with the WMAP5 value $\tau_{\rm CMB}=0.087\pm0.017$ \citep{dunkley2009}.  
Planck, which will have better polarisation sensitivity, should improve this significantly with Fisher matrix calculations predicting constraints at the level of $\sigma_\tau=0.005$ \citep[e.g.][]{colombo2008}, although this may degrade to $\sigma_\tau=0.01$ once foreground modeling is included \citep{tegmark2000}. 

Since observations suggest reionization is essentially complete by $z\approx6.5$, only a fraction of \tauCMB requires further explanation.  Assuming hydrogen and HeI are fully ionized by $z=6.5$ and that HeII reionization occurs at $z=3$ gives $\tau_{\rm CMB}=0.044$.  This leaves about half of the observed optical depth $\Delta\tau\approx0.043$ to be explained by ionization at higher redshifts.

In the instantaneous model used by the WMAP5 analysis \citep{dunkley2009}, the observed \tauCMB corresponds to a reionization redshift of $z_{\rm ri}=11\pm1.4$.  For more extended models of reionization, such as those of \citet{furlanetto2006}, full reionization occurs later, but with a significant tail to higher redshifts.  Since \tauCMB is an integral over the ionization history, the optical depth constraint essentially gives us a constraint on the duration of reionization or equivalently the redshift at which star formation began.  With much better data, one could imagine finding inconsistencies between \tauCMB and direct measurements of the reionization history indicating that other processes were responsible for modifying the ionization history at high redshift \citep{chen2004dm,mack2008}.  In this paper, we assume that all of \tauCMB arises from reionization by astrophysical sources and use the reported WMAP5 \tauCMB value.

In principle, CMB observations are sensitive to more than just \tauCMB since the shape of the reionization history will modify the polarization angular power spectrum on angular scales $\ell\lesssim50$ \citep{kaplinghat2003}.  Unfortunately, realistic reionization histories lead to only small modifications that are hard to detect with even cosmic variance limited observations \citep{holder2003}. It is therefore unlikely that CMB missions will provide much information on reionization beyond that contained in \tauCMB \citep{zaldarriaga2008}.

Turning now to astrophysical constraints, we begin with observations of the \lyf in quasar absorption spectra. These place constraints on the ionizing background produced by luminous sources.  The measured quantity is the mean transmittance of \lya flux $\langle F \rangle$ along the line of site to a quasar or equivalently the effective optical depth $\tau_{\rm eff}=-\log\langle F\rangle$  \citep{schaye2003,songaila2004,fan2006,fg2008b}.  With knowledge of the physical properties of the IGM, \tauEff can be connected to the background photoionization rate per hydrogen atom $\Gamma=\Gamma_{-12}\times 10^{-12}\,{\rm s^{-1}}$.  This quantity serves as a constraint on the sources of ionizing radiation that drive reionization.  

However, the conversion of a measured value of $\tau_{\rm eff}$ into \gammaI is a non-trivial process requiring the use of hydro-dynamical simulations of the IGM \citep{bolton2005, bolton2007} or the use of the fluctuating Gunn-Peterson approximation (FGPA) \citep{fan2006,fg2008}.  It is further plagued by our poor understanding of the relationship between overdensity $\Delta$ and temperature $T_{\rm IGM}$ in the IGM.  This is traditionally modelled at a power law $T=T_0\Delta^\beta$, the asymptotic form after reionization has taken place \citep{hui2003}.  However, recent simulations have shown that the situation at $z\sim3$ may be significantly more complicated as a result of HeII reionization \citep{bolton2008,hy2008,mcquinn2009} with there being some evidence for a temperature inversion.  At $z\gtrsim4$, there are currently no direct measurements of this relation \citep{cen2009}.

Since $\Gamma$ is a mostly local measurement, it primarily constrains the sources at the redshift at which it is measured. In this paper, we assume that our model of sources should consistently drive reionization and also match observations shortly after reionization has completed \citep{meiksin2005}.  It is important to remember that ionizing photons may come from quasars in addition to stars.  \citet{fg2008} show that, whereas the contribution from quasars drops rapidly at redshift $z\gtrsim4$ \citep{hopkins2007}, the contribution from galaxies remains fairly constant, suggesting that galaxies are the dominant source of ionizing photons during reionization.  Since it is these sources that we will be interested in constraining, we will consider only \gammaI constraints at $z\gtrsim4$.

\begin{table}
\caption{Constraints on \tauEff used in our analysis (taken from \citet{bolton2007} and based on data from \citet{schaye2003}, \citet{songaila2004}, and \citet{fan2006}).}
\begin{center}
\begin{tabular}{ccc}
$z\:$  & \tauEff & $\sigma_{\tau_{\rm eff}}$\\
\hline
3$\;$ & 0.362 & 0.035  \\
4$\;$  & 0.805 & 0.067 \\
5$\;$  & 2.07 & 0.27 \\
6$\;$  & $\gtrsim5.50$ & -\\
\end{tabular}
\end{center}
\label{tab:tau_bolton}
\end{table}%
The constraints on \tauEff that we use are shown in Table \ref{tab:tau_bolton} and taken from \citet[][henceforth \citetalias{bolton2007}]{bolton2007}.  Importantly, these contain invaluable data points at $z=5$ and $z=6$, while other authors have focussed on $z\lesssim4$.  The constraints from \citet{fg2008} (henceforth \citetalias{fg2008}), which utilise a larger quasar sample but extend only to $z=4.2$, are consistent with these, although the \citetalias{bolton2007} points are slightly lower at $z\sim4$ (probably as a result of not correcting for ``continuum bias". See \citetalias{fg2008} for discussion).  These two authors quote constraints on \gammaI that are systematically offset from one another, mostly as a result of different assumptions about the uncertain temperature-density relation.  We therefore choose to work from \tauEff and not \gammaI, so as to have control of the assumptions that are included.  Our model for converting \tauEff into \gammaI is detailed in \S\ref{sec:fgpa}.

As alluded to above, a knowledge of the temperature-density relationship of the IGM is an important ingredient in converting \tauEff observations into \gammaI constraints.  The IGM temperature can also potentially constrain when reionization took place.  After reionization, adiabatic cooling rapidly brings the IGM temperature-density relation to an asymptotic form \citep{hui2003}.  Observing an excess temperature over this might indicate that reionization had occurred recently, so that not enough time to reach the asymptote had elapsed.  Measurements of the IGM temperature come from \lya forest observations at $z\lesssim4$ from fitting to line widths \citep{schaye2000,ricotti2000,mcdonald2001} or from measurements of the small scale flux power spectrum \citep{zaldarriaga2001}.
The picture arising from these measurements is unclear and when the temperature-density relationship is parametrized as a power-law $T_{\rm IGM}=T_0\Delta^\beta$ the parameters are only loosely constrained in the range $T_0\approx(1-3)\times10^4\K$ and $\beta\approx0-0.8$, depending on the technique and data set.  Given the current level of the data and the importance of HeII reionization, we choose not to include temperature constraints in our analysis and only use these observations to set a reasonable prior on allowed values of $T_0$ and $\beta$ (see \citet{choudhury2005} for an alternative approach).  

Constraints on \gammaI may be converted into constraints on the rate per unit volume at which sources produce ionizing photons \Ndotns, given assumptions about the source spectrum, the mean free path of ionizing photons, and the distribution of absorbing systems.  These properties are poorly constrained by direct observation, as is discussed in more detail in \S\ref{sec:gamma2ndot}. 

Turning now to the future, more direct constraints on the ionization history may be possible with future 21 cm observations.  Experiments such as MWA, LOFAR, PAPER, and SKA will observe the 21 cm power spectrum $P_{21}(k)$ over a range of wavenumber $k$ in different redshift bins.  Analytic models and simulations suggest that the shape of $P_{21}(k)$
on scales $k\sim0.1-1\iMpc$ will provide good constraints on the bubble filling fraction $x_i(z)$ in each redshift bin.  Fisher matrix analyses \citep{bowman2006,mcquinn2006}
suggest that MWA will be able to place constraints on $x_H$ in a fully neutral IGM  at $z=8$ at the $\sigma_{x_i}=0.2$ level, while SKA should be capable of placing constraints of $\sigma_{x_i}=0.07$.  Adding in Planck CMB data to break degeneracies with cosmological parameters improves these constraints by a factor of $\sim2$. Similarly, by looking at the evolution of the amplitude and slope of $P_{21}$, \citet{lidz2007} find that MWA could constrain the midpoint of reionization with accuracy  $x_i=0.5\pm0.05$.  21 cm experiments may also give information about astrophysical parameters such as the minimum mass in which the first galaxies form \citep{barkana2008}.  

The success of these instruments is dependent upon reionization occurring within their sensitivity window.  These experiments are initially only likely to be sensitive to the power spectrum for frequencies above 100 MHz (although they are designed to cover frequencies down to 80 MHz), which corresponds to redshifts $z\lesssim13$.  At higher redshifts, synchrotron emission from the galaxy becomes extreme, terrestrial radio interference becomes more problematic, and the Earth's ionosphere is more of a challenge.  
It is therefore important to gauge what the ionization history is doing in this range of redshifts.

A second use of the 21 cm line comes from looking at the evolution of the global brightness temperature \citep{shaver1999}, rather its fluctuations.  Experiments such as EDGES \citep{bowman2007edges}, which look for a sharp transition in the 21 cm signal brought about by reionization essentially constrain $\ud T_b/\ud z$ and so $\ud x_i/\ud z$.  Beating experimental systematics down by a factor of 10 or so should enable these experiments to begin placing meaningful constraints on $x_i(z)$.  Again, it is important to ascertain the amplitude of the signal that these experiments may hope to see.

\section{Inference of ionization history} 
\label{sec:inference}

We next summarise our methodology for inferring the ionization history.  For a summary of Bayesian inference see e.g. \citet{mackay}.

We wish to attempt predictions for 21 cm observations given observational constraints from the \lyf and the CMB.  To do this we make use of Bayes theorem
\begin{equation}\label{bayes}
p(w | D,M)=\frac{p(D |w, M) p (w|M)}{p(D|M)},
\end{equation}
where $M$ is a model for reionization with parameters $w$, and $D$ is the combination of observational constraints on \Ndot and $\tau_{\rm CMB}$.  The evidence $p(D|M)$ provides an overall normalization for the posterior probability $p(w|D,M)$.  We must specify our prior $p(w|M)$ on the space of model parameters.  Where measurements are poor, we take flat priors over a specified range for each parameter, thus $p(w|M)={\rm const}$.  Our choice of prior should only be important if the data only weakly constrains the ionization history.  With these assumptions the posterior probability is proportional to the likelihood $p(D|w,M)$, which is straightforward to calculate. Note that we will make the simplifying assumption of Gaussian errors in the observational constraints on \tauEff and \tauCMB when calculating the likelihood, although this is by no means guaranteed.  

The first step of our inference procedure will be to use Eq. \eqref{bayes} and the data to constrain the model parameters for a set of different parametrizations.  Then, since each model provides a definite prediction for $x_i(z)$, we can calculate the probability distribution for $x_i$ at a given redshift from
\begin{equation}
p(x_i|M)=\int \ud w\, p(w | D,M) \delta[x_i(w|M)-x_i]
\end{equation}
This gives us a bound on the ionization history given our choice of the form of the source model.  The same procedure can be applied to calculate bounds on other quantities such as $\ud x_i/\ud z$ or the $S/N$ for a 21 cm experiment.  

The evolution of sources is likely to be complex and to be resistant to description by a small set of numbers.  However, by looking at a handful of models and seeing if the predictions are relatively consistent, we can still hope to obtain meaningful predictions.

\section{Results}
\label{sec:results}

\subsection{Constraints on \Ndot}
\label{sec:constraints}

We begin by mapping the constraints in \tauEff into constraints on \gammaI and \Ndotns.  This has been the subject of extensive work \citep{bolton2005,bolton2007,fg2008}, with each author making their own preferred choice of assumptions about uncertain astrophysical parameters.  In order to control the assumptions used  and to preserve the effect of the non-linear mapping of constraints, we perform our own analysis, but where possible we have checked that our model, which is described in detail in  Appendix \ref{sec:modeling}, reproduces the results of previous work.

\begin{table}
\caption{Fiducial parameter choices}
\begin{center}
\begin{tabular}{c|ccc}
Parameter & $\bar{x}$ ($x_{\rm low}$) & $\sigma_x$ ($x_{\rm high}$)& prior\\
\hline
 $T_0$ & $0.5\times10^4$K & $3.0\times10^4$K & uniform\\
$\beta$ & $0$ & $0.6$ & uniform\\
$\alpha_S$ & $1$ & $3$ & uniform\\
$\gamma$ & $1$ & $2$ & uniform\\
$\kappa$ & $1$ & $0.2$ & gaussian\\
$\sigma_8$ & 0.8 & 0.05 & gaussian \\
$\Omega_m$ & 0.3 & 0.04 & gaussian \\
$\Omega_b$ & 0.046 & 0.0005 & gaussian \\
$h$ & 0.7 & 0.04 & gaussian \\ 
\end{tabular}
\end{center}
\label{tab:model_prior}
\end{table}%
The main parameters in our model of reionization are the two parameters describing the temperature-density relation $T_0$ and $\beta$, the spectral index of the sources $\alpha_S$, and the power law index of absorbing systems $\gamma$.  We further allow for an uncertainty in the overall normalisation of the mean free path $\kappa$.  Alongside these astrophysical parameters, we allow for uncertainty in the normalisation of the matter power spectrum $\sigma_8$, the total matter density $\Omega_m$, the baryon density $\Omega_b$, and the Hubble constant $h$.

\begin{figure}
\begin{center}
\includegraphics[scale=0.4]{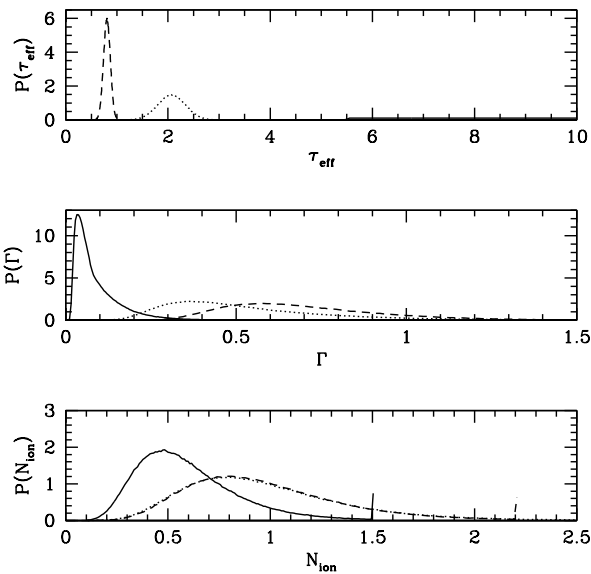}
\caption{Likelihoods for \tauEff (top panel), \gammaI (middle panel), and \Ndot (bottom panel, in units of $10^{51}{\,\rm s^{-1}\,Mpc^{-3}}$).  In each case, we plot the distribution for $z=4$ (dashed curves), $z=5$ (dotted curves), and $z=6$ (solid curves).}
\label{fig:gamdist_comp}
\end{center}
\end{figure}

Assuming the parameter priors in Table \ref{tab:model_prior}, we propagate uncertainties by randomly drawing specific realisations of the parameters from their respective distributions, calculating the resulting \gammaI and \Ndot and repeating in order to fully sample the final distributions.  Figure \ref{fig:gamdist_comp} shows the resulting uncertainty on \gammaI and \Ndotns.  Notice that in each case there is considerable uncertainty on each of the parameters.  

\begin{table}
\caption{Summary of constraints.  For \gammaI and \Ndot we show the most likely value and errors containing 68-\% and 95-\% confidence intervals.}
\begin{center}
\begin{tabular}{c|ccc}
Redshift & $\tau_{\rm eff}$ & \gammaI & \Ndot \\
\hline
4 & $0.805\pm0.067$ &$0.57^{+0.35\,(+0.71)}_{-0.10\,(-0.22)}$ & $0.80^{+0.53\,(+1.1)}_{-0.19\,(-0.40)}$\\
5 & $2.07\pm0.27$ & $0.36^{+0.39\,(+0.83)}_{-0.06\,(-0.16)}$& $0.80^{+0.53\,(+1.2)}_{-0.21\,(-0.42)}$\\
6 & $5.5-15$ &$0.03^{+0.11\,(+0.25)}_{-0.002\,(-0.02)}$ & $0.48^{+0.34\,(+0.75)}_{-0.12\,(-0.25)}$\\
\end{tabular}
\end{center}
\label{tab:constraints}
\end{table}%
Our constraints at each redshift are summarised in Table \ref{tab:constraints} where we list the most likely value as well as error bars containing 68-\% and 95-\% confidence intervals.  Besides the basic astrophysical parameters, we allow for an uncertainty in the overall normalisation of the mean free path and for uncertainty in the cosmological parameters.  A more detailed analysis would take into account the covariance between the observational uncertainties and these parameters, but that lies beyond the scope of our analysis. Note that the error bars are highly asymmetric, emphasising the need to consider the detailed probability distribution in our analysis.  The top panel of Figure \ref{fig:zeta_history} shows these error bars alongside those from \citetalias{bolton2007} and \citetalias{fg2008}.  Our error bars are deliberately quite conservative. 

Note that we treat the upper limit on \tauEff at $z=6$ as a uniform prior over a wide range of \tauEff values.  Increasing the upper limit here skews the constraint on \Ndot to lower values.  Unfortunately, this is at the edge of observational capability.  Improving these observations would go a long way towards constraining the evolution of the sources.

Note that the \Ndot constraints at $z=4$ and $z=5$ are essentially identical.  This is an indication that the errors from parameters other than \gammaI dominate the error budget.  Note that our constraints on \gammaI are somewhat different from those of \citetalias{fg2008} and \citetalias{bolton2007}  since these are very sensitive to the temperature prior used.  The overall normalization of the temperature is less important for the \Ndot constraints since shifting the temperature changes the normalisation of \gammaIns, but the scaling of $\lambda_{\rm mfp}$ is such as to compensate in the calculation of \Ndot (see Equation \eqref{gamma2ndot}). 

\subsection{Modeling ionizing sources} 
\label{sec:sources}

We next discuss our parametrization of the sources and consider two cases to explore the model dependence of our predictions.  Since we are focussed on behaviour at $z\gtrsim4$, we neglect the known contribution of quasars, which is expected to be small at these redshifts.  The arbitrary nature of our parametrization of \Ndot implicitly allows for a variety of contributing sources especially galaxies and quasars, but could also include X-ray ionizations from early mini-quasars \citep{ricotti2004} provided that these were unimportant for the \lyfns. 

The quantity \Ndot is affected by two ingredients: the star formation rate and the number of ionizing photons per baryon in stars that escapes to ionize in the IGM.  This motivates separating the two parts using the assumption that the star formation rate simply tracks the rate at which collapsed structures form.  This leads to
\begin{equation}
\dot{N}_{\rm ion}(z)=\zeta(z) n_H(0) \frac{\ud f_{\rm coll}(z)}{\ud t},
\end{equation}
where we have factored out the unknown number of ionizing photons per baryon, escape fraction, and star forming efficiency into a single ionizing efficiency $\zeta=N_{\rm UV} f_{\rm esc} f_{\star}$.  We calculate $f_{\rm coll}$ assuming a Press-Schecter mass function \citep{ps1974mfn} and using the minimum galaxy mass from atomic hydrogen line cooling.  Rather than attempt to model \Ndot directly, we can then model $\zeta(z)$ instead.  This has the advantage of separating out the expected rapid increase in the star formation rate, which results from the exponential increase of the collapse fraction. A physically motivated model is one that interpolates between two constant values
\begin{equation}
\zeta(z)=\zeta_0+\frac{(\zeta_1-\zeta_0)}{2}\left[\tanh\left(\frac{z-z_0}{\Delta z}\right)+1\right]
\end{equation}
This would result, for example, if an early period of Population III stars gave way to a later epoch of Population II star formation \citep{bromm2006}. 
For this parametrization, we expect the \Ndot constraints will fix the low $z$ amplitude of $\zeta$ while the optical depth constrains the high $z$ contribution from the second population.  Having four free parameters is really the minimum needed to allow for two populations of sources and so provide an interesting degree of flexibility in fitting the data.

It might also be natural to model \Ndot directly as a polynomial with some redshift at which star formation switches on, i.e. 
\begin{multline}
\dot{N}_{\rm ion}=N_0 A_{\rm ion}[1+N_1(z-z_0)+N_2(z-z_0)^2+N_3(z-z_0)^3]\\\times\Theta(z-z_{\rm max}),
\end{multline}
where the normalisation $A_{\rm ion}=10^{51} {\,\rm s^{-1}\,Mpc^{-3}}$, $z_0=4$, and we vary $z_{\rm max}$, $N_0$, $N_1$, and $N_2$ and adjust $N_3$ so that $\dot{N}_{\rm ion}(z_{\rm max})=0$.  This allows considerable flexibility to fit the data, allowing for many qualitatively different ionization histories.  In order to ensure that this parametrization does not lead to unphysically large values of \Ndot we impose the weak prior that $\dot{N}_{\rm ion}/A_{\rm ion}<10$ at all redshifts.  This only affects the results when we do not include the $z=6$ \lyf constraint.

Even this brief discussion illustrates the problem of a suitable parametrization.  With future observations, one could imagine more sophisticated modelling, but, as we shall show, current data is only sufficient to constrain two or three parameters making more detailed modelling premature.  

\begin{figure}
\begin{center}
\includegraphics[scale=0.4]{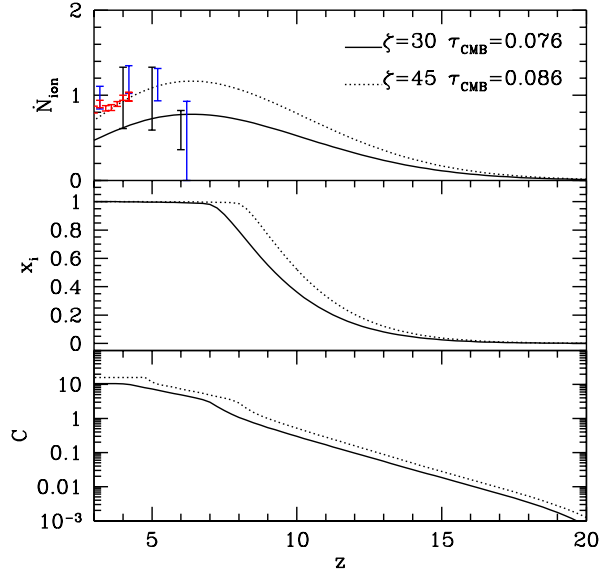}
\caption{Models with constant $\zeta(z)=30$ (solid curves) and $\zeta=45$ (dotted curves).  We show the redshift evolution of \Ndot (top panel, in units of $10^{51}{\,\rm s^{-1}\,Mpc^{-3}}$), $x_i$ (middle panel), and the clumping factor $C$ (bottom panel).  In the top panel, we show error bars on \Ndot calculated using the results of \citetalias{bolton2007} (blue) and \citetalias{fg2008} (red - statistical only) alongside our constraints (black, see \S\ref{sec:constraints}).}
\label{fig:zeta_history}
\end{center}
\end{figure}
To illustrate some of the general features of our modeling, we show in Figure \ref{fig:zeta_history} models with a constant value of $\zeta=30$ and $\zeta=45$.  This serves as a benchmark for illustrating some of the more generic features of our model.  The constant $\zeta$ leads to \Ndot tracing the rate of collapse of structure, which peaks at $z\approx 6$ in our model and drops going to lower redshifts where the contribution from quasars is expected to take over and dominate.  A lower value of $\zeta$ leads to a delay in the ionization history, so that reionization completes at redshifts different by $\Delta z\approx1$, and a lower Thomson optical depth to the CMB.  In both cases, recombinations play little role until reionization is well underway.

\subsection{Ionization history}
\label{sec:history}

We now use these constraints on \Ndot and \tauCMB to constrain our model parameters.  To explore the effect of CMB data, we consider combinations of the \lyf with WMAP3, WMAP5, and Planck.  Although the central value of the observed \tauCMB appears constant between data sets, there is still the possibility of systematic errors at the level of $\sigma_{\tau_{\rm CMB}}=0.01$ in the WMAP5 data.  Relaxing the constraint on \tauCMB gives a sense of how these systematic errors might change our results.  We also consider constraints from WMAP5 in combination with only the $z=4$ and $z=5$ \lyf data, throwing out the $z=6$ constraint.  The $z=6$ \lyf is subject to considerable theoretical uncertainty since it requires extrapolating from the IGM properties observed at $z=4$ the most.  Moreover, in late reionization models $z=6$ may occur close enough to overlap that spatial fluctuations in the temperature-density relationship \citep{furlanetto2009temp} or ionizing background \citep{furlanetto2009ion} may complicate the connection between \tauEff and \Ndotns.

We calculate the likelihood for the parameters by uniform sampling of the four dimensional parameter space.  In each case, we assume a uniform prior on the source parameters.  Although numerically somewhat inefficient this makes later analysis more straightforward.

In addition to the CMB and \lyf constraints, we make a cut throwing out all models with $x_i(z=6)<0.99$.  That the IGM is ionized to at least this level is well established by \lyf observations and imposing this constraint ensures that HII regions have percolated placing us in the range of validity of our \lyf modelling.

\subsubsection{$\zeta$ parametrization}
\label{sec:twostep}

\begin{figure}
\begin{center}
\includegraphics[scale=0.4]{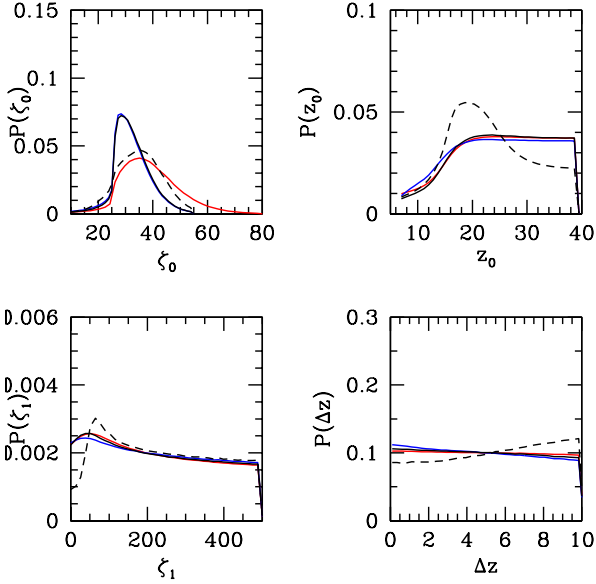}
\caption{Marginalised PDF for the $\zeta$ parametrization. Different combinations of data are plotted: \lyf + WMAP5 (black), \lyf (excluding $z=6$) + WMAP5 (red), \lyf + WMAP3 (blue), and \lyf + Planck (black dashed).}
\label{fig:twostep_pdf4}
\end{center}
\end{figure}
Our physically motivated $\zeta$ parametrization naturally describes two epochs of star formation with differing efficiencies.  Figure \ref{fig:twostep_pdf4} shows the marginalised probability distribution function (PDF) for the four model parameters.  We see that the \lyf constraints on \Ndot lead to a clear preference for $\zeta_0$ centered around $\zeta_0\approx30$ consistent with the ionizing efficiency expected for Pop II stars \citep[see e.g.][]{furlanetto2006}.  In order to satisfy the \tauCMB constraint however a population of more efficient sources at higher redshifts is required.  This could arise, for example, from evolution in the escape fraction with redshift or from the contribution of a different population of sources at high redshift e.g. Population III stars.  The model prefers that this transition occur at $z\gtrsim15$, but there is little constraint on exactly when this transition needs to take place, since a higher transition redshift can be compensated for with a higher $\zeta_1$ leading to a more extreme early burst of ionization.  No preference for the width of the transition $\Delta z$ is shown.  Including Planck type \tauCMB constraints significantly improves constraints on $\zeta_0$, $\zeta_1$, and $z_0$ since it reduces the freedom to change the high redshift behaviour of the model.

\begin{figure}
\begin{center}
\includegraphics[scale=0.4]{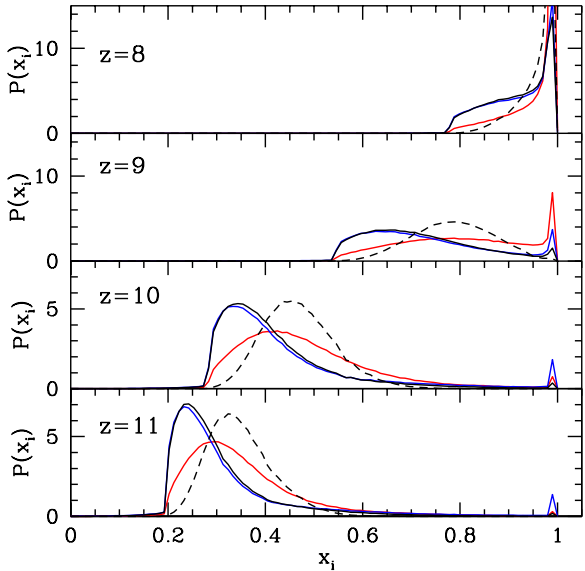}
\caption{Distribution of $x_i$ at redshifts $z=8$, 9, 10, and 11 for the $\zeta$ parametrization.  Same curve styles as for Figure \ref{fig:twostep_pdf4}. }
\label{fig:twostep_xi}
\end{center}
\end{figure}
In Figure \ref{fig:twostep_xi} we show the marginalised distribution of $x_i$ in four redshift bins of interest to upcoming 21 cm experiments.  Although there is considerable uncertainty in $x_i$ at each redshift the model does lead to clearly preferred ranges.  Including the $z=6$ \lyf point pushes the distribution to slightly lower values of $x_i$ since it pushes the normalisation of $\zeta$ downwards.  Within this parametrization, the IGM is highly ionized by $z=8$ (at least to $x_i\gtrsim0.8$).  Note that since $0\le x_i\le1$ there is a ``pile up" effect leading to a peak at $x_i=1$, where models leading to early reionization all contribute. 

\subsubsection{\Ndot parametrization}
\label{sec:ndot}

\begin{figure}
\begin{center}
\includegraphics[scale=0.4]{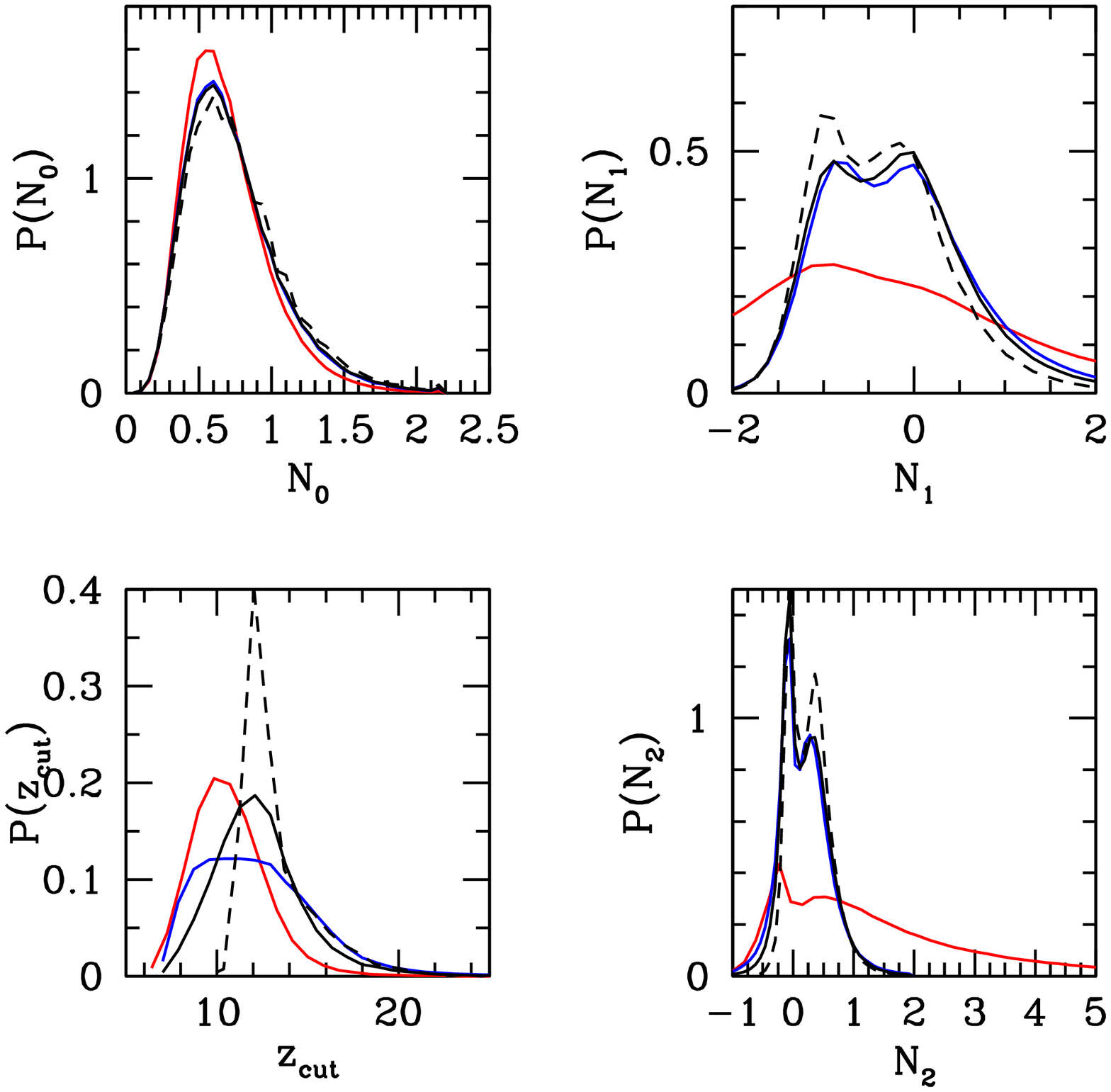}
\caption{Marginalised PDF for the \Ndot parametrization. Different combinations of data are plotted: \lyf + WMAP5 (black), \lyf (excluding $z=6$) + WMAP5 (red), \lyf + WMAP3 (blue), and \lyf + Planck (black dashed).}
\label{fig:ndot_pdf4}
\end{center}
\end{figure}
We now compare results from the \Ndot parametrization, analysed in the same as way as in \S\ref{sec:twostep}.  Figure \ref{fig:ndot_pdf4} shows constraints on the model parameters.  These display a bimodal distribution that arises because of the exclusion of models that do not ionize the Universe early enough. 

\begin{figure}
\begin{center}
\includegraphics[scale=0.4]{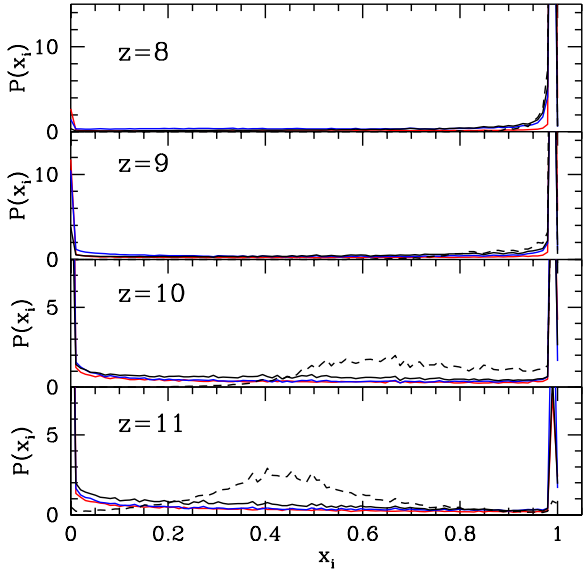}
\caption{Distribution of $x_i$ at redshifts $z=8$, 9, 10, and 11 for the \Ndot parametrization.   Same curve styles as for Figure \ref{fig:ndot_pdf4}. }
\label{fig:ndot_xi}
\end{center}
\end{figure}

 This model has significantly more flexibility than the $\zeta$ parametrization, which shows itself in the distribution of $x_i$ seen in Figure \ref{fig:ndot_xi}.  The distributions are very broad at $z=10$ and $z=11$ indicating that the data does a poor job in constraining possible histories within this parametrization.  It is important to note that even with this increased flexibility we still see that the IGM is largely ionized by $z=8$.  Planck level \tauCMB constraints lead to better localised distributions.  

\subsubsection{Comparison}
\label{sec:comparison}

\begin{figure}
\begin{center}
\includegraphics[scale=0.4]{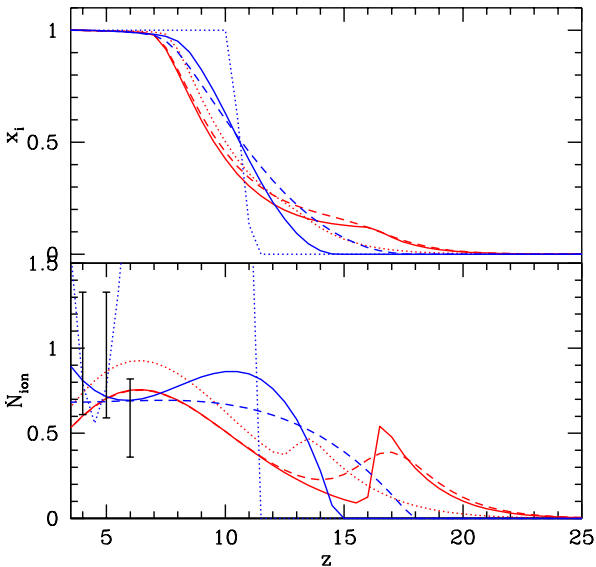}
\caption{Redshift evolution of $x_i$ (top panel) and \Ndot (bottom panel, in units of $10^{51}{\,\rm s^{-1}\,Mpc^{-3}}$) for the best fitting history for the $\zeta$ (red curves) and \Ndot (blue curves) parametrizations.  Different combinations of data are plotted: \lyf + WMAP5 (solid), \lyf (excluding $z=6$) + WMAP5 (dotted), and \lyf + Planck (dashed).  Error bars show the constraints on \Ndot from Table \ref{tab:constraints}.}
\label{fig:best_history}
\end{center}
\end{figure}
In Figure \ref{fig:best_history} we give a sense of the best fitting histories produced by the two different parametrizations and combinations of data.  There is clearly considerable spread in the sorts of history that are consistent with observations. 

\begin{figure}
\begin{center}
\includegraphics[scale=0.4]{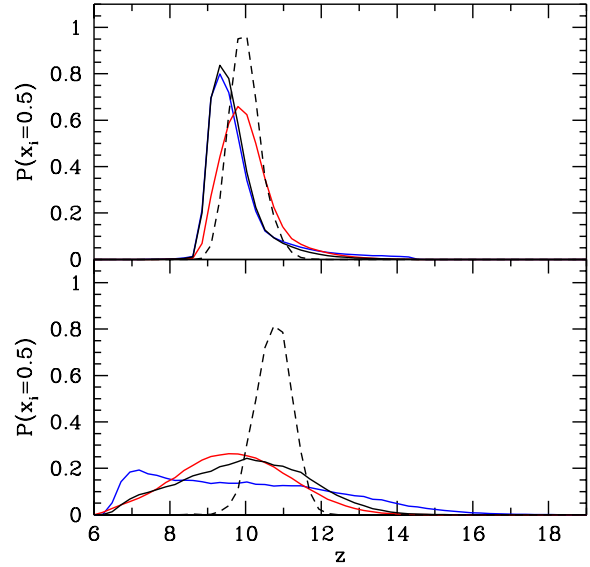}
\caption{Marginalised redshift distribution of $x_i(z)=0.5$ for the $\zeta$ (top panel) and \Ndot (bottom panel) parametrizations.  Same line conventions as for Figure \ref{fig:ndot_pdf4}. }
\label{fig:full_z_pdf}
\end{center}
\end{figure}

Two natural milestones in the reionization of the Universe are the point at which $x_i(z)=0.5$ and when reionization completes.  This first point, $x_i(z)=0.5$, is thought to lead to distinctive signatures in the 21 cm signal \citep{lidz2007}.  In Figure \ref{fig:full_z_pdf}, we plot the PDF for finding $x_i=0.5$ at a given redshift.  Although there are differences in the spread of uncertainty between our two parametrizations, we see that in each panel the peak of the distribution lies around $z=9-11$.  Existing data has a relatively strong preference for the Universe to be half ionized by $z=9$.  There is also a marked tail of the distribution at redshifts greater than this.

\begin{figure}
\begin{center}
\includegraphics[scale=0.4]{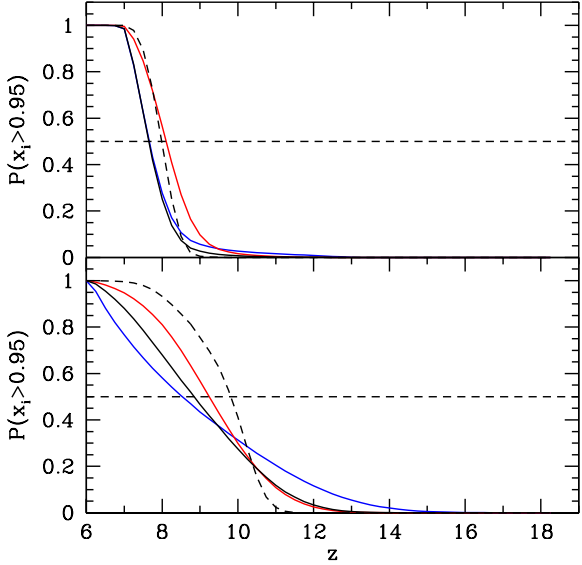}
\caption{Probability at a given redshift that $x_i>0.95$ for the $\zeta$  (top panel) and \Ndot (bottom panel) parametrizations.  Same line conventions as for Figure \ref{fig:ndot_pdf4}. }
\label{fig:final_z_pdf}
\end{center}
\end{figure}

The second milestone, the end of reionization, is a somewhat slippery moment to define.  One might define it as the point when bubbles percolate, so that $x_i=1$, or define it in terms of the point where the mean free path of photons grows to the horizon size, so that every gas element sees all sources.  Moreover making predictions for the final few percent of the ionized fraction depends sensitively on the mean free path of the photons and the clumping. For the purposes of planning 21 cm instruments, the important moment is when the filling fraction of neutral regions drops to a few percent, so that the signal strength becomes small.  In Figure \ref{fig:final_z_pdf}, we plot the probability that $x_i>0.95$ at a given redshift.  This clearly indicates that, in both parametrizations, reionization is preferred to be mostly complete by $z=8$.  It is important to note that, while the predictions for the midpoint of reionization are driven almost entirely by the CMB, the \lyf provides very useful information in determining when reionization completed.

It is clear from the preceding results that the \Ndot parametrization has much greater flexibility in fitting the \lyf data and so leads to more conservative results.  The $\zeta$ parametrizations is quite restrictive in that the low redshift behaviour is very well constrained by the \lya forest.  As a result, much of the tightness of the constraints in the $\zeta$ parametrization is driven by the form of the parametrization itself.  This highlights the importance of choosing parametrizations capable of representing a wide variety of reionization histories with only a few parameters.  

\subsection{21 cm fluctuations and observations}
\label{sec:21cm}
\subsubsection{Estimating the 21 cm power spectrum}

\begin{figure}
\begin{center}
\includegraphics[scale=0.4]{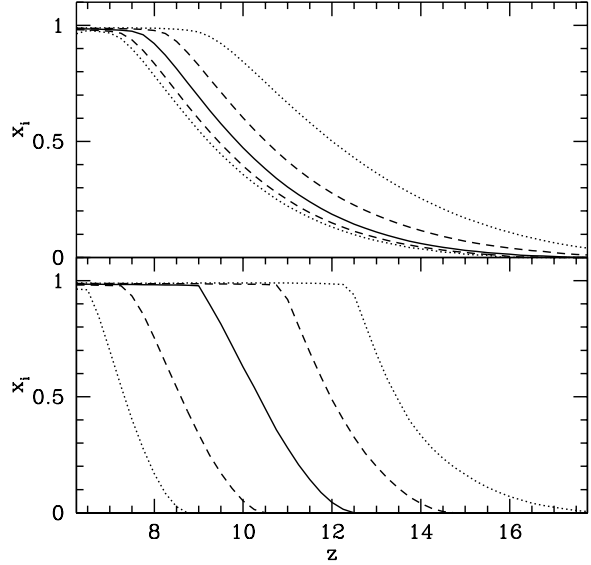}
\caption{Confidence intervals at the 95\% (dotted curves) and 68\% (dashed curves) levels and median history (solid curve) for the ionization history for the $\zeta$ (top panel) and \Ndot (bottom panel) parametrizations using \lyf and WMAP5.  }
\label{fig:xi_contour}
\end{center}
\end{figure}

We now turn to making predictions for 21 cm experiments.  Of particular interest is the ionization history.  In the previous section, we calculated the marginalised PDF of $x_i$ in separate redshift bins.  We use the same calculation to calculate constraints on $x_i$ as a function of $z$ using the cumulative probability to establish confidence intervals for the model.   Figure \ref{fig:xi_contour} shows 68-\% and 95-\% confidence intervals for the ionization history calculated in our two models.  It should not be surprising from our previous discussion that the constraints on $x_i(z)$ are significantly less tight in the more flexible \Ndot parametrization.  This is a strong indication of how parametrization dependent our results are (note that using a power law for $\zeta$ leads to similar constraints as a power law in \Ndotns).  Regardless of the parametrization, the Universe is likely to have been fully ionized by $z=8$, $p(x_i(z=8)>0.99)\gtrsim0.5$.

Although having a constraint on the ionization history is useful, it is advantageous to map this into the expected amplitude of the 21 cm brightness temperature power spectrum $P_{21}$.  In order to achieve this mapping, we need two pieces of information: the evolution of the mean brightness temperature and a mapping between $x_i(z)$ and the amplitude of the fluctuations on a given scale.  Predicting the evolution of the mean brightness temperature requires, in addition to a knowledge of $x_i(z)$, information about the IGM temperature and the \lya background responsible for coupling the spin and kinetic temperatures.  There is considerable uncertainty in predicting these quantities \citep[see e.g.][]{furlanetto2006, pritchard2007xray,pritchard2008}, so we will instead naively assume that the gas has been heated to $T_k\gg T_{\rm CMB}$ and that $T_S=T_k$ as a result of strong Wouthysen-Field coupling.  With these assumptions, we have
\begin{equation}
T_b=27 \,x_{\rm HI}\left(\frac{\Omega_b h^2}{0.023}\right)\left(\frac{0.15}{\Omega_mh^2}\frac{1+z}{10}\right)^{1/2}\,{\rm mK}.
\end{equation}
To map a given $x_i(z)$ into the desired $P_{21}(k,z)$, we assume the validity of the analytic model of \citet{fzh2004} to calculate the power spectrum of the ionized component $P_{xx}$.  This model, based on the excursion set formalism, has been shown to provide a good match to numerical simulations of reionization \citep{zahn2007,santos2007}.  Moreover it has been shown that the spectrum of fluctuations can be well modelled given only $x_i$; it is surprisingly robust to other astrophysical parameters once $x_i$ has been fixed \citep{mcquinn2007}.  We numerically build a lookup table specifying $P_{21}$ as a function of $z$ and $x_i$, which is then used to map our constraints on $x_i$ into constraints on $P_{21}(k)$, where we fix $k=0.1\iMpc$ a scale roughly in the center of the range accessible to the first generation of 21 cm experiments. 

\begin{figure}
\begin{center}
\includegraphics[scale=0.4]{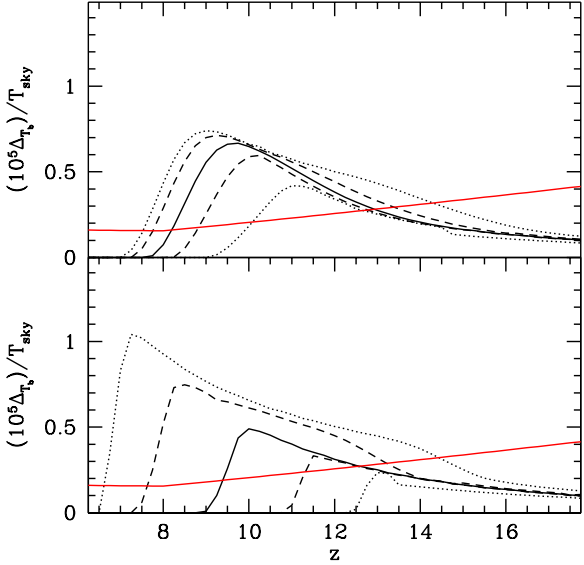}
\caption{Confidence intervals at the 95\% (dotted curves) and 68\% (dashed curves) levels and median history (solid curve) for $\Delta_{T_b}/T_{\rm sky}$ for the $\zeta$ (top panel) and \Ndot (bottom panel) parametrizations using \lyf and WMAP5.  Solid red curves indicate the thermal noise for MWA.}
\label{fig:fluc_contour}
\end{center}
\end{figure}

The sensitivity of the MWA to the 21cm power spectrum will be limited by thermal noise in the antennae rather than by survey volume. At the relevant frequencies the thermal noise will be dominated by the galactic foreground, which is observed to have a temperature $T_{\rm sky}\approx0.9(1+z)^{2.6}{\,\rm K}$. The thermal noise for the observations therefore rises steeply at high redshift, implying that better signal to noise will be obtained at lower redshifts. A useful outcome of our analysis is a prediction of the redshift where the MWA might be most likely to achieve the highest SN detection. The likelihood for a given redshift is proportional to the amplitude of the PS divided by the square of the temperature of the sky. Thus, consideration of the system noise weights the sensitivity of the instruments to lower redshift ranges.  Our results from this exercise are shown in Figure \ref{fig:fluc_contour}, where we plot the power per logarithmic interval $\Delta_{T_b}=\sqrt{k^3P_{21}/2\pi^2}$ divided by $T_{\rm sky}$ and for clarity in the scales have multiplied the resulting S/N by $10^5$. As expected the curves cut off at low $z$ as the gas becomes fully ionized and all the curves tend towards the same asymptote at high $z$ where the gas is fully neutral.  We find that 21 cm experiments are likely to achieve the highest signal-to-noise ratio in the range $z=9-10$.  

As an example of a 21 cm experiment, in Figure \ref{fig:fluc_contour}, we plot the expected thermal noise for MWA assuming the configuration described in \citet{mcquinn2006}, an integration of 2000 hours on two places on the sky and a bandwidth of 6 MHz.  We have allowed the collecting area of the dipoles to scale as $\lambda^2$, except above $z=8$ where the area has been capped to reflect geometric shadowing within the antennae tiles.  This clearly indicates that there is a good chance of MWA detecting the 21 cm signal over a wide range of redshifts.

\subsubsection{The global 21 cm signature}

\begin{figure}
\begin{center}
\includegraphics[scale=0.4]{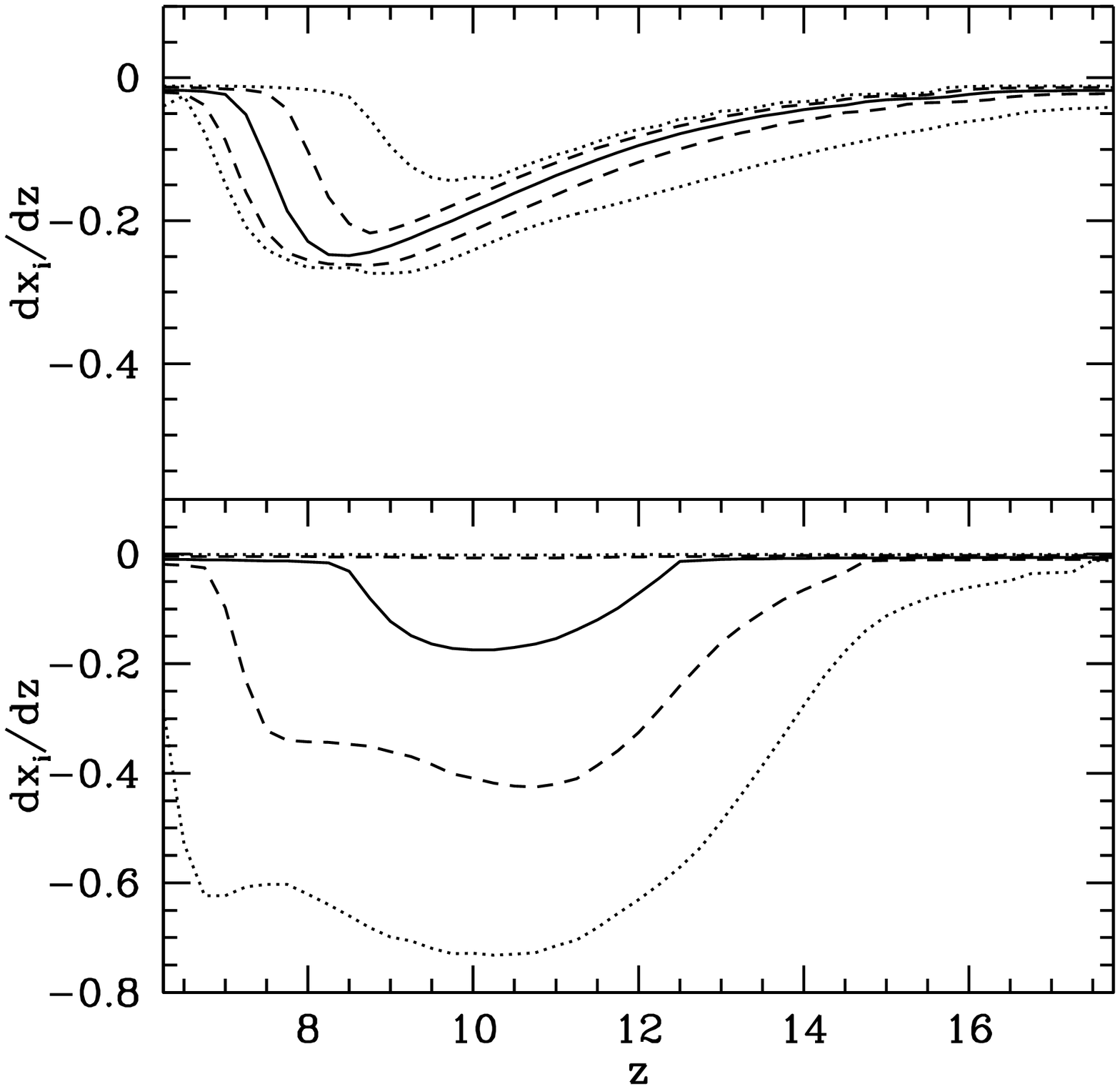}
\caption{Confidence intervals at the 95\% (dotted curves) and 68\% (dashed curves) levels and median history (solid curve) for $\ud x_i/\ud z$ for the $\zeta$ (top panel) and \Ndot (bottom panel) parametrizations using \lyf and WMAP5.}
\label{fig:dxdz_contour}
\end{center}
\end{figure}

Experiments sensitive to the global 21 cm signal will be most sensitive to sharp features in $x_i(z)$ where the derivative of $\ud x_i/\ud z$ is large, since smooth modes are removed with the foregrounds.  Converting the reionization history into the expected signal is difficult, since it depends upon both the foregrounds and the instrumental response, which modifies the shape of the foregrounds from a simple power-law.  In Figure \ref{fig:dxdz_contour}, we show the confidence intervals on $\ud x_i/\ud z$.  Since, in most of our histories, reionization takes place over $\Delta z\gtrsim2$ the derivative seldom becomes large.  Since our reionization histories are monotonic the derivative is always negative.  More detailed calculation, including the effects of gas heating and \lya coupling, would show more complicated behaviour.

Since 21cm observations are able to measure the ionization fraction $x_i$ during the central parts of reionization (Lidz et al.~2007), it is interesting to ask how such a measurement would aid our knowledge of reionization during other stages of the process.

\subsubsection{The impact of a 21 cm measurement}

\begin{figure}
\begin{center}
\includegraphics[scale=0.4]{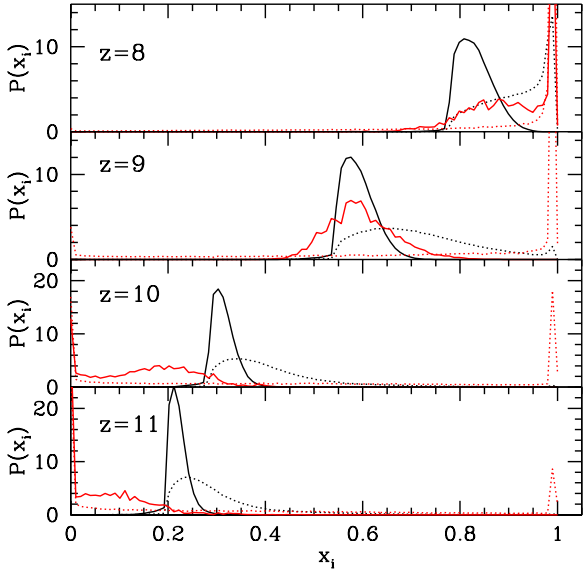}
\caption{Distribution of $x_i$ at redshifts $z=8$, 9, 10, and 11 when 21 cm measurements are included.   In each panel, we plot the distribution of the $\zeta$ (black) and \Ndot (red) parametrizations with (solid curves) and without (dotted curves) a 21 cm measurement of $x_i(z=9.5)=0.5\pm0.05$.}
\label{fig:tocm_xi}
\end{center}
\end{figure}

In Figure \ref{fig:tocm_xi}, we show the marginalised PDF for $x_i$ in several redshift bins given a measurement of $x_i(z=9.5)=0.5\pm0.05$.  Here we chose a fiducial value consistent with that of our previous analysis and take an uncertainty broadly consistent with that quoted in the analysis of \citep{lidz2007} for the MWA.  We see that adding a direct measurement of $x_i$ improves the constraints on the evolution of $x_i$ considerably.  This feeds into the constraints on the model parameters and conceivably into our knowledge of the underlying astrophysical sources.  

\subsubsection{Estimating \tauCMB from 21 cm measurements}

Finally, we consider a more speculative point.  From the point of view of constraining fundamental cosmological parameters, such as the tilt and running of the inflationary power spectrum, the optical depth to last scattering is a nuisance parameter.  If \tauCMB could be independently measured to high accuracy, so that it was in effect a known quantity, then it could be removed from the uncertainty in CMB parameter estimates.  At least in principle, 21 cm experiments, by placing direct constraints on $x_i$ in many redshift bins, could constrain \tauCMB directly.  Whether this is possible in practice will be determined by how much of the ionization history where $x_i>0$ is accessible to 21 cm observations.  Given that CMB constraints will approach $\sigma_{\tau_{\rm CMB}}\approx0.005$ most of the ionization history must be mapped for 21 cm constraints to be competitive.

\begin{figure}
\begin{center}
\includegraphics[scale=0.4]{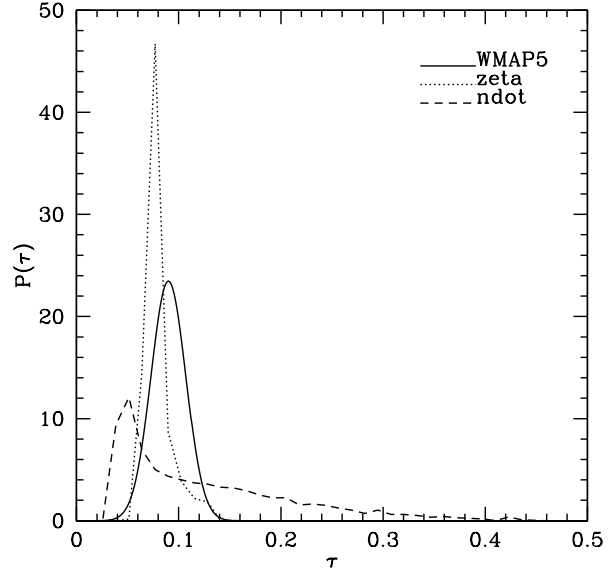}
\caption{Distribution of \tauCMB given 21 cm measurements of $x_i(z=8)\gtrsim0.8$ and $x_i(z=9)=0.5\pm0.05$ in combination with \lyf data for the $\zeta$ (dotted curve) and \Ndot (dashed curve) parametrizations.  For comparison, we plot the WMAP5 constraint (solid curve) assuming Gaussian errors.}
\label{fig:taudist}
\end{center}
\end{figure}
Figure \ref{fig:taudist} shows the errors on \tauCMB achieved within our modeling by adding two 21 cm measurements of $x_i(z=8)>0.8$ and $x_i(z=9)=0.5\pm0.05$, similar to what might be achieved by first generation experiments, to the \lyf data.  Clearly, the distributions are not competitive with the WMAP5 constraint unless the source evolution is quite constrained.  Nonetheless constraints on \tauCMB from the directly measured ionization history may provide a useful consistency check with the results from CMB experiments.

\section{Conclusions}
\label{sec:conclude}

In this paper, we have developed a framework for combining existing constraints on reionization from CMB and \lyf observations, and used them to place bounds on the reionization history.  Modeling of reionization is very uncertain, both in terms of the sources of radiation (their evolution, spectrum, clustering, etc.), and in terms of their interaction with the neutral IGM. Rather than employ a semi-analytic model, with all the necessary assumptions \citep[e.g.][]{choudhury2005}, we have therefore chosen a more general approach which utilises arbitrary forms for the evolution of ionizing emissivity. This approach avoids theoretical prejudice regarding the model for reionization at the greatest possible level.  The framework that we have used here can easily be extended to incorporate more general parametrizations of the sources as improved data makes that worthwhile.

The resulting analytic model for reionization and the \lyf is described in Appendix \ref{sec:modeling}.  This approach remains model dependent. However the level of dependance has been quantified by considering two different parametrizations for the sources. These give consistent predictions on $x_i(z)$, albeit with differing levels of precision.  Nevertheless, we are able to draw some relatively robust conclusions from our analysis.  The main conclusion is that current observations suggest reionization was largely complete by $z=8$ with the Universe likely to have been half ionized between $z=9-10$.  This result has important implications for upcoming 21 cm experiments. In particular we have shown that the signal-to-noise ratio for these experiments is most likely to be at its maximum in the range $z=9-10$. Thus current observables of reionization can be used to guide the first attempts at 21cm observations, which will be very expensive in terms of observing time, and cover only a fraction of the reionization era at a time.  We also demonstrate that following a 21cm derived measurement of the ionization fraction midway through the reionization era, the constraints on the entire history will be greatly enhanced.  Finally, we speculate that future 21 cm experiments might begin to constrain \tauCMB sufficiently to provide an independent consistency check on CMB results.

\section*{Acknowledgements}

JRP would like to thank Jamie Bolton, Claude-Andre Faucher-Giguere, Steven Furlanetto, and Adam Lidz for comments on an earlier draft and MPA for their hospitality while part of this work was completed.
JRP is supported by NASA through Hubble Fellowship grant HST-HF-01211.01-A
awarded by the Space Telescope Science Institute, which is operated by the
Association of Universities for Research in Astronomy, Inc., for NASA,
under contract NAS 5-26555. AL acknowledges funding from NSF grant AST-0907890 and from the NASA Lunar Science
 Institute (via Cooperative Agreement NNA09DB30A).  JSBW is supported by the Australian Research Council.


\begin{thebibliography}{}

\bibitem[\protect\citeauthoryear{{Barkana}}{{Barkana}}{2008}]{barkana2008}
{Barkana} R.,  2008, ArXiv e-prints, 0806.2333

\bibitem[\protect\citeauthoryear{{Barkana} \& {Loeb}}{{Barkana} \&
  {Loeb}}{2001}]{barkana2001}
{Barkana} R.,  {Loeb} A.,  2001, \physrep, 349, 125

\bibitem[\protect\citeauthoryear{{Barkana} \& {Loeb}}{{Barkana} \&
  {Loeb}}{2002}]{barkana2002}
{Barkana} R.,  {Loeb} A.,  2002, \apj, 578, 1

\bibitem[\protect\citeauthoryear{{Becker}, {Rauch} \& {Sargent}}{{Becker}
  et~al.}{2009}]{becker2009}
{Becker} G.~D.,  {Rauch} M.,    {Sargent} W.~L.~W.,  2009, \apj, 698, 1010

\bibitem[\protect\citeauthoryear{{Becker} et~al.,}{{Becker}
  et~al.}{2001}]{becker2001}
{Becker} R.~H.,  et~al., 2001, \aj, 122, 2850

\bibitem[\protect\citeauthoryear{{Bolton} \& {Becker}}{{Bolton} \&
  {Becker}}{2009}]{bolton2009}
{Bolton} J.~S.,  {Becker} G.~D.,  2009, ArXiv e-prints, 0906.2861

\bibitem[\protect\citeauthoryear{{Bolton} \& {Haehnelt}}{{Bolton} \&
  {Haehnelt}}{2007}]{bolton2007}
{Bolton} J.~S.,  {Haehnelt} M.~G.,  2007, \mnras, 382, 325

\bibitem[\protect\citeauthoryear{{Bolton}, {Haehnelt}, {Viel} \&
  {Springel}}{{Bolton} et~al.}{2005}]{bolton2005}
{Bolton} J.~S.,  {Haehnelt} M.~G.,  {Viel} M.,    {Springel} V.,  2005, \mnras,
  357, 1178

\bibitem[\protect\citeauthoryear{{Bolton}, {Viel}, {Kim}, {Haehnelt} \&
  {Carswell}}{{Bolton} et~al.}{2008}]{bolton2008}
{Bolton} J.~S.,  {Viel} M.,  {Kim} T.-S.,  {Haehnelt} M.~G.,    {Carswell}
  R.~F.,  2008, \mnras, 386, 1131

\bibitem[\protect\citeauthoryear{{Bouwens}, {Illingworth}, {Thompson} \&
  {Franx}}{{Bouwens} et~al.}{2005}]{bouwens2005}
{Bouwens} R.~J.,  {Illingworth} G.~D.,  {Thompson} R.~I.,    {Franx} M.,  2005,
  \apjl, 624, L5

\bibitem[\protect\citeauthoryear{{Bowman}, {Morales} \& {Hewitt}}{{Bowman}
  et~al.}{2006}]{bowman2006}
{Bowman} J.~D.,  {Morales} M.~F.,    {Hewitt} J.~N.,  2006, \apj, 638, 20

\bibitem[\protect\citeauthoryear{{Bowman}, {Rogers} \& {Hewitt}}{{Bowman}
  et~al.}{2007}]{bowman2007edges}
{Bowman} J.~D.,  {Rogers} A.~E.~E.,    {Hewitt} J.~N.,  2007, ArXiv e-prints,
  0710.2541

\bibitem[\protect\citeauthoryear{{Bromm}, {Kudritzki} \& {Loeb}}{{Bromm}
  et~al.}{2001}]{bromm2001}
{Bromm} V.,  {Kudritzki} R.~P.,    {Loeb} A.,  2001, \apj, 552, 464

\bibitem[\protect\citeauthoryear{{Bromm} \& {Loeb}}{{Bromm} \&
  {Loeb}}{2006}]{bromm2006}
{Bromm} V.,  {Loeb} A.,  2006, \apj, 642, 382

\bibitem[\protect\citeauthoryear{{Cen}, {McDonald}, {Trac} \& {Loeb}}{{Cen}
  et~al.}{2009}]{cen2009}
{Cen} R.,  {McDonald} P.,  {Trac} H.,    {Loeb} A.,  2009, ArXiv e-prints,
  0907.0735

\bibitem[\protect\citeauthoryear{{Chen} \& {Kamionkowski}}{{Chen} \&
  {Kamionkowski}}{2004}]{chen2004dm}
{Chen} X.,  {Kamionkowski} M.,  2004, \prd, 70, 043502

\bibitem[\protect\citeauthoryear{{Choudhury} \& {Ferrara}}{{Choudhury} \&
  {Ferrara}}{2005}]{choudhury2005}
{Choudhury} T.~R.,  {Ferrara} A.,  2005, \mnras, 361, 577

\bibitem[\protect\citeauthoryear{{Choudhury} \& {Ferrara}}{{Choudhury} \&
  {Ferrara}}{2006}]{choudhury2006}
{Choudhury} T.~R.,  {Ferrara} A.,  2006, \mnras, 371, L55

\bibitem[\protect\citeauthoryear{{Colombo}, {Pierpaoli} \&
  {Pritchard}}{{Colombo} et~al.}{2008}]{colombo2008}
{Colombo} L.~P.~L.,  {Pierpaoli} E.,    {Pritchard} J.~R.,  2008, ArXiv
  e-prints, 0811.2622

\bibitem[\protect\citeauthoryear{{Dunkley} et~al.,}{{Dunkley}
  et~al.}{2009}]{dunkley2009}
{Dunkley} J.,  et~al., 2009, \apjs, 180, 306

\bibitem[\protect\citeauthoryear{{Fan}, {Strauss}, {Becker} et~al.,}{{Fan}
  et~al.}{2006}]{fan2006}
{Fan} X.,  {Strauss} M.~A.,  {Becker} R.~H.,    et~al., 2006, \aj, 132, 117

\bibitem[\protect\citeauthoryear{{Faucher-Gigu{\`e}re}, {Lidz}, {Hernquist} \&
  {Zaldarriaga}}{{Faucher-Gigu{\`e}re} et~al.}{2008a}]{fg2008}
{Faucher-Gigu{\`e}re} C.-A.,  {Lidz} A.,  {Hernquist} L.,    {Zaldarriaga} M.,
  2008a, ArXiv e-prints, 0806.0372

\bibitem[\protect\citeauthoryear{{Faucher-Gigu{\`e}re}, {Lidz}, {Hernquist} \&
  {Zaldarriaga}}{{Faucher-Gigu{\`e}re} et~al.}{2008b}]{fg2008c}
{Faucher-Gigu{\`e}re} C.-A.,  {Lidz} A.,  {Hernquist} L.,    {Zaldarriaga} M.,
  2008b, \apj, 688, 85

\bibitem[\protect\citeauthoryear{{Faucher-Gigu{\`e}re}, {Prochaska}, {Lidz},
  {Hernquist} \& {Zaldarriaga}}{{Faucher-Gigu{\`e}re} et~al.}{2008}]{fg2008b}
{Faucher-Gigu{\`e}re} C.-A.,  {Prochaska} J.~X.,  {Lidz} A.,  {Hernquist} L.,
   {Zaldarriaga} M.,  2008, \apj, 681, 831

\bibitem[\protect\citeauthoryear{{Furlanetto}}{{Furlanetto}}{2006}]{furlanetto%
2006}
{Furlanetto} S.~R.,  2006, \mnras, 371, 867

\bibitem[\protect\citeauthoryear{{Furlanetto} \& {Mesinger}}{{Furlanetto} \&
  {Mesinger}}{2009}]{furlanetto2009ion}
{Furlanetto} S.~R.,  {Mesinger} A.,  2009, \mnras, 394, 1667

\bibitem[\protect\citeauthoryear{{Furlanetto} \& {Oh}}{{Furlanetto} \&
  {Oh}}{2005}]{furlanetto2005}
{Furlanetto} S.~R.,  {Oh} S.~P.,  2005, \mnras, 363, 1031

\bibitem[\protect\citeauthoryear{{Furlanetto} \& {Oh}}{{Furlanetto} \&
  {Oh}}{2009}]{furlanetto2009temp}
{Furlanetto} S.~R.,  {Oh} S.~P.,  2009, \apj, 701, 94

\bibitem[\protect\citeauthoryear{{Furlanetto}, {Zaldarriaga} \&
  {Hernquist}}{{Furlanetto} et~al.}{2004}]{fzh2004}
{Furlanetto} S.~R.,  {Zaldarriaga} M.,    {Hernquist} L.,  2004, \apj, 613, 1

\bibitem[\protect\citeauthoryear{{Gunn} \& {Peterson}}{{Gunn} \&
  {Peterson}}{1965}]{gp1965}
{Gunn} J.~E.,  {Peterson} B.~A.,  1965, \apj, 142, 1633

\bibitem[\protect\citeauthoryear{{Holder}, {Haiman}, {Kaplinghat} \&
  {Knox}}{{Holder} et~al.}{2003}]{holder2003}
{Holder} G.~P.,  {Haiman} Z.,  {Kaplinghat} M.,    {Knox} L.,  2003, \apj, 595,
  13

\bibitem[\protect\citeauthoryear{{Hopkins}, {Richards} \&
  {Hernquist}}{{Hopkins} et~al.}{2007}]{hopkins2007}
{Hopkins} P.~F.,  {Richards} G.~T.,    {Hernquist} L.,  2007, \apj, 654, 731

\bibitem[\protect\citeauthoryear{{Hui} \& {Haiman}}{{Hui} \&
  {Haiman}}{2003}]{hui2003}
{Hui} L.,  {Haiman} Z.,  2003, \apj, 596, 9

\bibitem[\protect\citeauthoryear{{Kaplinghat}, {Chu}, {Haiman}, {Holder},
  {Knox} \& {Skordis}}{{Kaplinghat} et~al.}{2003}]{kaplinghat2003}
{Kaplinghat} M.,  {Chu} M.,  {Haiman} Z.,  {Holder} G.~P.,  {Knox} L.,
  {Skordis} C.,  2003, \apj, 583, 24

\bibitem[\protect\citeauthoryear{{Komatsu}, {Dunkley}, {Nolta}
  et~al.,}{{Komatsu} et~al.}{2009}]{komatsu2009}
{Komatsu} E.,  {Dunkley} J.,  {Nolta} M.~R.,    et~al., 2009, \apjs, 180, 330

\bibitem[\protect\citeauthoryear{{Leitherer}, {Schaerer}, {Goldader}
  et~al.,}{{Leitherer} et~al.}{1999}]{leitherer1999}
{Leitherer} C.,  {Schaerer} D.,  {Goldader} J.~D.,    et~al., 1999, \apjs, 123,
  3

\bibitem[\protect\citeauthoryear{{Lidz}, {Zahn}, {McQuinn}, {Zaldarriaga} \&
  {Hernquist}}{{Lidz} et~al.}{2007}]{lidz2007}
{Lidz} A.,  {Zahn} O.,  {McQuinn} M.,  {Zaldarriaga} M.,    {Hernquist} L.,
  2007, ArXiv e-prints, 0711.4373

\bibitem[\protect\citeauthoryear{{Mack} \& {Wesley}}{{Mack} \&
  {Wesley}}{2008}]{mack2008}
{Mack} K.~J.,  {Wesley} D.~H.,  2008, ArXiv e-prints, 0805.1531

\bibitem[\protect\citeauthoryear{MacKay}{MacKay}{2003}]{mackay}
MacKay D.,  2003, {Information Theory, Inference, and Learning Algorithms}.
Cambridge University Press

\bibitem[\protect\citeauthoryear{{Malhotra} \& {Rhoads}}{{Malhotra} \&
  {Rhoads}}{2004}]{malhotra2004}
{Malhotra} S.,  {Rhoads} J.~E.,  2004, \apjl, 617, L5

\bibitem[\protect\citeauthoryear{{McDonald}, {Miralda-Escud{\'e}}, {Rauch},
  {Sargent}, {Barlow} \& {Cen}}{{McDonald} et~al.}{2001}]{mcdonald2001}
{McDonald} P.,  {Miralda-Escud{\'e}} J.,  {Rauch} M.,  {Sargent} W.~L.~W.,
  {Barlow} T.~A.,    {Cen} R.,  2001, \apj, 562, 52

\bibitem[\protect\citeauthoryear{{McQuinn}, {Lidz}, {Zahn}, {Dutta},
  {Hernquist} \& {Zaldarriaga}}{{McQuinn} et~al.}{2007}]{mcquinn2007}
{McQuinn} M.,  {Lidz} A.,  {Zahn} O.,  {Dutta} S.,  {Hernquist} L.,
  {Zaldarriaga} M.,  2007, \mnras, 377, 1043

\bibitem[\protect\citeauthoryear{{McQuinn}, {Lidz}, {Zaldarriaga}, {Hernquist}
  \& {Dutta}}{{McQuinn} et~al.}{2008}]{mcquinn2008grb}
{McQuinn} M.,  {Lidz} A.,  {Zaldarriaga} M.,  {Hernquist} L.,    {Dutta} S.,
  2008, \mnras, 388, 1101

\bibitem[\protect\citeauthoryear{{McQuinn}, {Lidz}, {Zaldarriaga}, {Hernquist},
  {Hopkins}, {Dutta} \& {Faucher-Gigu{\`e}re}}{{McQuinn}
  et~al.}{2009}]{mcquinn2009}
{McQuinn} M.,  {Lidz} A.,  {Zaldarriaga} M.,  {Hernquist} L.,  {Hopkins} P.~F.,
   {Dutta} S.,    {Faucher-Gigu{\`e}re} C.-A.,  2009, \apj, 694, 842

\bibitem[\protect\citeauthoryear{{McQuinn}, {Zahn}, {Zaldarriaga}, {Hernquist}
  \& {Furlanetto}}{{McQuinn} et~al.}{2006}]{mcquinn2006}
{McQuinn} M.,  {Zahn} O.,  {Zaldarriaga} M.,  {Hernquist} L.,    {Furlanetto}
  S.~R.,  2006, \apj, 653, 815

\bibitem[\protect\citeauthoryear{{Meiksin}}{{Meiksin}}{2005}]{meiksin2005}
{Meiksin} A.,  2005, \mnras, 356, 596

\bibitem[\protect\citeauthoryear{{Miralda-Escud{\'e}}}{{Miralda-Escud{\'e}}}{2%
003}]{me2003}
{Miralda-Escud{\'e}} J.,  2003, \apj, 597, 66

\bibitem[\protect\citeauthoryear{{Miralda-Escud{\'e}}, {Haehnelt} \&
  {Rees}}{{Miralda-Escud{\'e}} et~al.}{2000}]{me2000}
{Miralda-Escud{\'e}} J.,  {Haehnelt} M.,    {Rees} M.~J.,  2000, \apj, 530, 1

\bibitem[\protect\citeauthoryear{{Misawa}, {Tytler}, {Iye}, {Kirkman},
  {Suzuki}, {Lubin} \& {Kashikawa}}{{Misawa} et~al.}{2007}]{misawa2007}
{Misawa} T.,  {Tytler} D.,  {Iye} M.,  {Kirkman} D.,  {Suzuki} N.,  {Lubin} D.,
     {Kashikawa} N.,  2007, \aj, 134, 1634

\bibitem[\protect\citeauthoryear{{Parsons}, {Backer}, {Bradley}
  et~al.,}{{Parsons} et~al.}{2009}]{parsons2009}
{Parsons} A.~R.,  {Backer} D.~C.,  {Bradley} R.~F.,    et~al., 2009, ArXiv
  e-prints, 0904.2334

\bibitem[\protect\citeauthoryear{{Pawlik}, {Schaye} \& {van
  Scherpenzeel}}{{Pawlik} et~al.}{2009}]{pawlik2009}
{Pawlik} A.~H.,  {Schaye} J.,    {van Scherpenzeel} E.,  2009, \mnras, 394,
  1812

\bibitem[\protect\citeauthoryear{{Press} \& {Schechter}}{{Press} \&
  {Schechter}}{1974}]{ps1974mfn}
{Press} W.~H.,  {Schechter} P.,  1974, \apj, 187, 425

\bibitem[\protect\citeauthoryear{{Pritchard} \& {Furlanetto}}{{Pritchard} \&
  {Furlanetto}}{2007}]{pritchard2007xray}
{Pritchard} J.~R.,  {Furlanetto} S.~R.,  2007, \mnras, 376, 1680

\bibitem[\protect\citeauthoryear{{Pritchard} \& {Loeb}}{{Pritchard} \&
  {Loeb}}{2008}]{pritchard2008}
{Pritchard} J.~R.,  {Loeb} A.,  2008, \prd, 78, 103511

\bibitem[\protect\citeauthoryear{{Ricotti}, {Gnedin} \& {Shull}}{{Ricotti}
  et~al.}{2000}]{ricotti2000}
{Ricotti} M.,  {Gnedin} N.~Y.,    {Shull} J.~M.,  2000, \apj, 534, 41

\bibitem[\protect\citeauthoryear{{Ricotti} \& {Ostriker}}{{Ricotti} \&
  {Ostriker}}{2004}]{ricotti2004}
{Ricotti} M.,  {Ostriker} J.~P.,  2004, \mnras, 352, 547

\bibitem[\protect\citeauthoryear{{Santos}, {Amblard}, {Pritchard}, {Trac},
  {Cen} \& {Cooray}}{{Santos} et~al.}{2007}]{santos2007}
{Santos} M.~G.,  {Amblard} A.,  {Pritchard} J.,  {Trac} H.,  {Cen} R.,
  {Cooray} A.,  2007, ArXiv e-prints, 0708.2424

\bibitem[\protect\citeauthoryear{{Schaye}}{{Schaye}}{2001}]{schaye2001}
{Schaye} J.,  2001, \apj, 559, 507

\bibitem[\protect\citeauthoryear{{Schaye}, {Aguirre}, {Kim}, {Theuns}, {Rauch}
  \& {Sargent}}{{Schaye} et~al.}{2003}]{schaye2003}
{Schaye} J.,  {Aguirre} A.,  {Kim} T.-S.,  {Theuns} T.,  {Rauch} M.,
  {Sargent} W.~L.~W.,  2003, \apj, 596, 768

\bibitem[\protect\citeauthoryear{{Schaye}, {Theuns}, {Rauch}, {Efstathiou} \&
  {Sargent}}{{Schaye} et~al.}{2000}]{schaye2000}
{Schaye} J.,  {Theuns} T.,  {Rauch} M.,  {Efstathiou} G.,    {Sargent}
  W.~L.~W.,  2000, \mnras, 318, 817

\bibitem[\protect\citeauthoryear{{Scheuer}}{{Scheuer}}{1965}]{scheuer1965}
{Scheuer} P.~A.~G.,  1965, \nat, 207, 963

\bibitem[\protect\citeauthoryear{{Shaver}, {Windhorst}, {Madau} \& {de
  Bruyn}}{{Shaver} et~al.}{1999}]{shaver1999}
{Shaver} P.~A.,  {Windhorst} R.~A.,  {Madau} P.,    {de Bruyn} A.~G.,  1999,
  \aap, 345, 380

\bibitem[\protect\citeauthoryear{{Songaila}}{{Songaila}}{2004}]{songaila2004}
{Songaila} A.,  2004, \aj, 127, 2598

\bibitem[\protect\citeauthoryear{{Stark}, {Ellis}, {Richard}, {Kneib}, {Smith}
  \& {Santos}}{{Stark} et~al.}{2007}]{stark2007}
{Stark} D.~P.,  {Ellis} R.~S.,  {Richard} J.,  {Kneib} J.-P.,  {Smith} G.~P.,
   {Santos} M.~R.,  2007, \apj, 663, 10

\bibitem[\protect\citeauthoryear{{Storrie-Lombardi}, {McMahon}, {Irwin} \&
  {Hazard}}{{Storrie-Lombardi} et~al.}{1994}]{storrie1994}
{Storrie-Lombardi} L.~J.,  {McMahon} R.~G.,  {Irwin} M.~J.,    {Hazard} C.,
  1994, \apjl, 427, L13

\bibitem[\protect\citeauthoryear{{Tanvir} et~al.,}{{Tanvir}
  et~al.}{2009}]{tanvir2009}
{Tanvir} N.~R.,  et~al., 2009, ArXiv e-prints, 0906.1577

\bibitem[\protect\citeauthoryear{{Tegmark}, {Eisenstein}, {Hu} \& {de
  Oliveira-Costa}}{{Tegmark} et~al.}{2000}]{tegmark2000}
{Tegmark} M.,  {Eisenstein} D.~J.,  {Hu} W.,    {de Oliveira-Costa} A.,  2000,
  \apj, 530, 133

\bibitem[\protect\citeauthoryear{{Trac} \& {Cen}}{{Trac} \&
  {Cen}}{2007}]{trac2007}
{Trac} H.,  {Cen} R.,  2007, \apj, 671, 1

\bibitem[\protect\citeauthoryear{{Trac}, {Cen} \& {Loeb}}{{Trac}
  et~al.}{2008}]{hy2008}
{Trac} H.,  {Cen} R.,    {Loeb} A.,  2008, \apjl, 689, L81

\bibitem[\protect\citeauthoryear{{Trac} \& {Gnedin}}{{Trac} \&
  {Gnedin}}{2009}]{trac2009}
{Trac} H.,  {Gnedin} N.~Y.,  2009, ArXiv e-prints, 0906.4348

\bibitem[\protect\citeauthoryear{{Weinberg} \& {et al.}}{{Weinberg} \& {et
  al.}}{1999}]{weinberg1999}
{Weinberg} D.,  {et al.} 1999, in {Banday} A.~J.,  {Sheth} R.~K.,   {da Costa}
  L.~N.,  eds, Evolution of Large Scale Structure : From Recombination to
  Garching {Cosmological tests with the Ly-{$\alpha$} forest (invited review)}.
p.~346

\bibitem[\protect\citeauthoryear{{Zahn}, {Lidz}, {McQuinn}, {Dutta},
  {Hernquist}, {Zaldarriaga} \& {Furlanetto}}{{Zahn} et~al.}{2007}]{zahn2007}
{Zahn} O.,  {Lidz} A.,  {McQuinn} M.,  {Dutta} S.,  {Hernquist} L.,
  {Zaldarriaga} M.,    {Furlanetto} S.~R.,  2007, \apj, 654, 12

\bibitem[\protect\citeauthoryear{{Zaldarriaga}, {Colombo}, {Komatsu}
  et~al.,}{{Zaldarriaga} et~al.}{2008}]{zaldarriaga2008}
{Zaldarriaga} M.,  {Colombo} L.,  {Komatsu} E.,    et~al., 2008, ArXiv
  e-prints, 0811.3918

\bibitem[\protect\citeauthoryear{{Zaldarriaga}, {Hui} \&
  {Tegmark}}{{Zaldarriaga} et~al.}{2001}]{zaldarriaga2001}
{Zaldarriaga} M.,  {Hui} L.,    {Tegmark} M.,  2001, \apj, 557, 519

\end{thebibliography}

\appendix

\section{Modeling reionization} 
\label{sec:modeling}

In this Appendix, we describe the details of our model for reionization and the framework that we use to connect different observables.  We begin with our model for the \lya forest, which leads naturally into predictions for the evolution of the IGM neutral fraction and thus the CMB optical depth.  This model makes a number of simplifying assumptions and should not be considered a true substitute to detailed numerical simulation.  It does however provide a reasonable fit to existing observations and enables us to evaluate the large region of parameter space necessary for our analysis.  

\subsection{Mapping \tauEff to \gammaI}   
 \label{sec:fgpa}
  
We wish to convert the observed values of the \lya effective optical depth, which is defined via the fractional transmittance
\begin{equation}
\tau_{\rm eff}\equiv -\log[\langle F\rangle(z)],
\end{equation}
into values for \gammaI.  To achieve this we make use of the fluctuating Gunn-Peterson approximation \citep{weinberg1999, fg2008c}.  This provides a self-consistent model for the \lya forest, although it leads to parameter scalings that only approximate more detailed numerical simulations \citep{bolton2005}.

Neglecting redshift-space distortions
\begin{equation}
F=\exp(-\tau),
\end{equation}
  where, assuming photo-ionization equilibrium, the optical depth to \lya photons is given by
  \begin{equation}
\tau=\frac{\pi e^2 f_{\rm Ly\alpha}}{m_e \nu_{\rm Ly\alpha}}\frac{1}{H(z)}\frac{R(T)n_{\rm HII}n_e}{\Gamma},
\end{equation}
in terms of the recombination rate $R(T)$ and the metagalactic ionization rate per hydrogen atom $\Gamma=\Gamma_{-12}\times 10^{-12}{\,\rm s^{-1}}$.  For case A recombination, $R(T)=R_0 T^{-0.7}$, with $R_0=4.2\times10^{-13}{\,\rm cm^3\,s^{-1}}/(10^4{\,\rm K})^{-0.7}$.  In order to apply this, we assume a power law relationship between the IGM temperature and density
\begin{equation}
T=T_0\Delta^\beta,
\end{equation}
with $\Delta=1+\delta$.  The unknown quantities $T_0$ and $\beta$ represent a significant uncertainty in our interpretation of the \lya forest.  We will consider different values in our analysis.

The observations of \tauEff come from averaging over the spectra of many quasars.  We therefore average the transmittance (specified by the above equations as a function of $\Delta$) over a probability density function (PDF) for $\Delta$ to get
\begin{equation}
\langle F\rangle(z)=\int_0^{\infty}\ud\Delta\, P(\Delta;z)\exp(-\tau).
\end{equation}

\begin{figure}
\begin{center}
\includegraphics[scale=0.4]{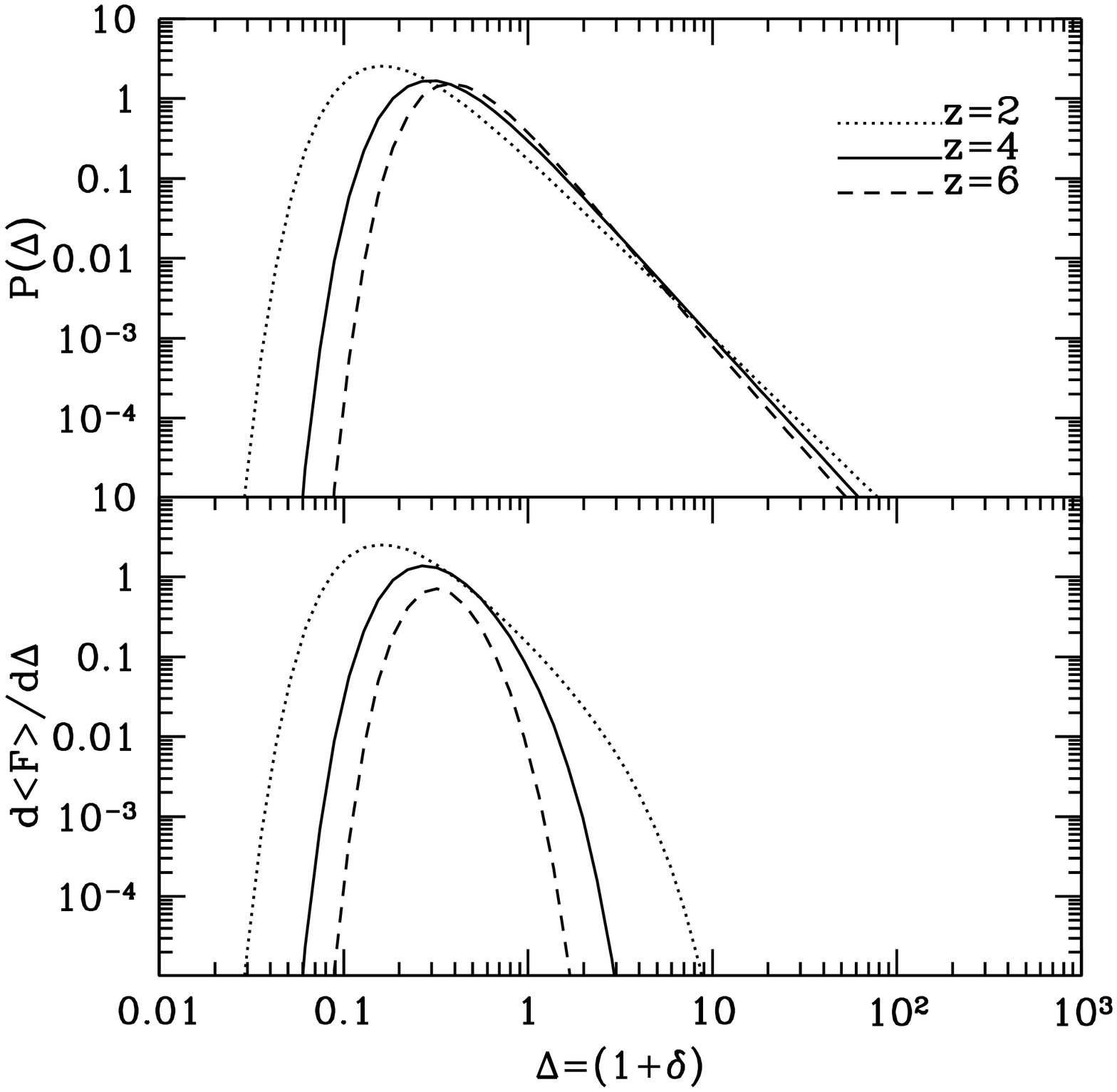}
\caption{{\em Top panel: }The PDF for $\Delta=1+\delta$ from MHR00.  Note the lognormal shape with a long power law tail at high $\Delta$.  This PDF has been shown to fail for $\Delta\gtrsim10$ and is also suspect at very low $\Delta$.  {\em Bottom panel: }The kernel for calculating the mean transmittance $\langle F\rangle$.  It peaks at $\Delta\sim0.3$, corresponding to slightly underdense regions.  The calculation will be affected by errors in the PDF at low $\Delta$.}
\label{fig:lya_pdf}
\end{center}
\end{figure}

We make use of the analytic $P(\Delta)$ from \citet{me2000}.  Several recent papers \citep{pawlik2009, bolton2009} have shown that this PDF fails to match numerical simulations at high densities where the PDF has a power law tail.  The integral for $\langle F\rangle$ obtains most of its contribution from low-density regions (see Figure \ref{fig:lya_pdf}), with non-zero transmittance.  At lower $z$, where the mean density is lower, larger overdensities begin to contribute. Using the analytic PDF of \citep{me2000} rather than the more accurate numerical fit of \citet{bolton2009} makes an $\sim10\%$ difference in the conversion of \tauEff into \gammaIns.  In summary, given the above equations, a PDF for $\Delta$, and knowledge of $T_0$ and $\beta$, we can map a given value of \tauEff to the corresponding value of \gammaIns.

\subsection{Mapping from \gammaI to \Ndot}
\label{sec:gamma2ndot}
We next seek to relate \gammaI to the rate at which sources produce ionizing photons \Ndotns.  The comoving ionization rate is
\begin{equation}
\Gamma=(1+z)^2\int_{\nu_{0}}^{\infty}\ud\nu\,\epsilon_\nu \lambda_\nu \sigma_\nu,
\end{equation}
in terms of the emissivity of sources $\epsilon_\nu\sim\nu^{-\alpha_S-1}$, the mean free path $\lambda_\nu$, and the ionization cross section $\sigma_\nu\sim\nu^{-3}$.  The mean free path here introduces another source of considerable uncertainty.  The frequency dependence of $\lambda_\nu$ depends upon the distribution of column density of absorbing systems, so that for a power law distribution of column densities $f(\tau)\sim\tau^{-\gamma}$ the mean free path scales as $\lambda_\nu=(\nu/\nu_{0})^{3(\gamma-1)}\lambda_{\rm mfp}(\nu_0)$, where $\lambda_0$ is the mean free path at the Lyman limit $\nu_0$.  Taking these frequency scalings leads to the relation 
\begin{eqnarray}\label{gamma2ndot}
\dot{N}_{\rm ion}&=& 10^{51.2}\, \Gamma_{-12}\left(\frac{\alpha_S}{3}\right)^{-1}\left(\frac{\alpha_S+3(2-\gamma)}{6}\right)\\
&&\times\left(\frac{\lambda_{\rm mfp}(\nu_0)}{\rm 40\, Mpc}\right)^{-1}\left(\frac{1+z}{7}\right)^{-2}{\rm s^{-1}\,Mpc^{-3}}.
\end{eqnarray}
Here $\alpha_S$ is the spectral index of the sources just above the HI ionization threshold, since the frequency dependence of the cross-section causes the contribution of higher energy photons to drop rapidly. This spectral index enters as a source of uncertainty in the overall normalization of the inferred \Ndotns and we allow for a range between $\alpha_S=1$ and $\alpha_S=3$ spanning the range expected from Population III stars \citep{bromm2001} with hard spectra through to softer galaxy spectra \citep[e.g][]{leitherer1999}.

In the post-overlap reionization model of \citet{me2000}, the mean free path of a photon is given by
\begin{equation}\label{mfp}
\lambda_{\rm mfp}=\lambda_0(1+z)[1-F_V(\Delta<\Delta_i)]^{-2/3}.
\end{equation}
Here $F_V(\Delta_i)=\int^{\Delta_i}\ud\Delta\,P(\Delta)$ is the fraction of gas by volume contained in regions with density below a critical density for ionization $\Delta<\Delta_i$.  In order to calculate $\Delta_i$ and $\lambda_0$ we must model the \lya forest in more detail, which we do following the arguments of \citet{furlanetto2005}.  We may associate the column density of a Lyman limit system to the critical density by assuming that the characteristic size is the local Jeans length \citep{schaye2001} and that photoionization equilibrium holds.  Thus, assuming that $N_{\rm HI}\approx x_{\rm HI} \Delta \bar{n}_H L_J$ gives 
\begin{equation}
N_{\rm HI}=3.3\times10^{17}{\rm\,cm^{-2}}\left(\frac{\Delta}{100}\right)^{3/2}T_4^{-0.26}\Gamma_{-12}^{-1}\left(\frac{1+z}{7}\right)^{9/2}.
\end{equation}
Setting $N_{\rm HI}(\Delta)\approx1/\sigma_0=1.6\times10^{17}{\rm\,cm^{-2}}$ gives the critical overdensity for a self-shielding clump as
\begin{equation}
\Delta_i\approx49.5\left(\frac{T}{10^4\K}\right)^{0.13}\left(\frac{1+z}{7}\right)^{-3}\Gamma_{-12}^{2/3}.
\end{equation}
We will be focusing on $z\gtrsim4$ where $\Delta_i\lesssim150$, as such our density PDF should be valid.  At higher densities, we might expect the details of star formation to become important leading to an extra source of uncertainty.

This prescription implies a distribution of column densities in the \lya forest 
\begin{equation}
\frac{\ud^2 N}{\ud N_{\rm HI}\,\ud z}=\frac{(1+z)^2}{H(z)}\Omega_b\frac{\ud\Delta}{\ud N_{\rm HI}}\Delta P(\Delta)\frac{3c(1-Y)}{8\pi G m_H}x_{\rm HI} N_{\rm HI}^{-1}.
\end{equation}

This model provides a reasonable description of the distribution of column densities of absorbing systems and gives us a way of calculating the unknown absorption probability per unit length $\lambda_0^{-1}$.  An estimate of the spacing between Lyman limit systems is
\begin{equation}
\lambda_{\rm LLS}=\frac{c H^{-1}}{(1+z)}\left(\frac{\ud N_{\rm LLS}}{\ud z}\right)^{-1}.
\end{equation}

The mean free path should be comparable to this spacing, but we need to account for the distribution of systems with differing column density to incorporate the cumulative effect of lower column-density systems.  For a photon at the hydrogen ionization absorption edge, this gives \citep{me2003}
\begin{equation}\label{opacity}
\frac{1}{\lambda_{\rm mfp}}=\frac{\int_0^\infty\ud\tau\, \tau^{-\gamma}(1-e^{-\tau})}{\int_1^\infty\ud\tau\, \tau^{-\gamma}\lambda_{\rm LLS}},
\end{equation}
where we have defined $\tau=N_{\rm HI}/(1.6\times10^{16}\,{\rm cm^{-2}})$.
Using the value of $\gamma$ corresponding to $z=3$ gives $\lambda_{\rm mfp}\approx\lambda_{\rm LLS}/2$.  Hence, there is a redshift dependent numerical factor of about 2 connecting $\lambda_0$ with the mean separation between Lyman limit systems.  Although this model agrees well with the results of \citet{me2000},  we should be wary of the possibility of uncertainties in the normalisation arising from the ambiguity of using the Jeans' length as the characteristic size of the objects.  In our calculations, we will therefore allow for a normalisation offset $\kappa=\lambda_{\rm mfp}/\lambda_{\rm mfp,model}$.  Setting out the calculation in this way facilitates comparison with observations of the number of Lyman limit systems per unit redshift \citep{storrie1994} and the distribution of systems of different column density \citep{misawa2007}.  Using this model and observational input, we may convert observations of \tauEff into constraints on \Ndotns.  The main sources of uncertainty arise from $\alpha_S$ and $\lambda_0$ and from the temperature-density relation.

\subsection{Evolution of the neutral fraction}
\label{sec:neutralfrac}

Given $\dot{N}_{\rm ion}$, we calculate the HII region filling fraction $Q_{\rm HII}$ using
\begin{equation}
\frac{\ud Q_{\rm HII}}{\ud t}=\frac{\dot{N}_{\rm ion}}{n_H(0)}-Q_{\rm HII} C\, n_H(z)\alpha_A(T).
\end{equation}
where $C\equiv\langle n_{\rm HII}^2\rangle/\langle n_{\rm HII}\rangle^2$ is the clumping of ionized hydrogen.  Note that we assume case-A recombination, this is appropriate since the majority of recombinations occur on the edge of self-shielding dense clumps which will trap the ionizing photons produced by recombination direct to the ground state preventing them from contributing to ionization in the IGM.

Besides the sources, the crucial uncertainty in the evolution of the neutral fraction lies in the dependence of the recombination rate upon the clumping $C$.  This quantity is expected to be redshift dependent and vary significantly from $C\lesssim 1$ at high redshifts if low density regions are ionized first, to $C\gtrsim 4$ at low redshift, where the remaining neutral gas is located in overdense regions.  This variation may have significant effect on the reionization history, since a high clumping factor may delay the completion of reionization by $\Delta z\sim2$.  It is therefore important that we explore a range of values for the clumping and allow for its redshift evolution.  Note that the clumping will generally be a function of both redshift and neutral fraction.

In the model of \citetalias{me2000}, the recombination rate in an inhomogeneous Universe is related to the uniform recombination rate $R_u$ by
\begin{equation}
R=R_u\int^{\Delta_i} \ud\Delta P_V(\Delta) \Delta^2\equiv C R_u.
\end{equation}
By itself, this formalism would underestimate the effective clumping, since reionization proceeds via the percolation of ionized bubbles, which are produced by highly biased sources.  This may be accounted for by calculating the recombination rate inside each bubble and averaging over the distribution of bubble sizes.  We do this following \citet{furlanetto2005}.  Accounting for the bubble distribution increases the clumping factor by a factor of $\sim4$.  This model predicts clumping factors that rise rapidly from values $C\ll1$ early in reionization to values $C\sim10$ as reionization comes close to completion and shows behaviour that is qualitatively similar to that found in simulations \citep[e.g.][]{trac2007}.  Figure \ref{fig:zeta_history} shows an example of the evolution of $C$ from this calculation.

One aspect that we ignore is the photoionization of mini-halos, which could consume UV photons and thus delay reionization by $\Delta z\approx1$ \citep{barkana2002}.  The clumping could also be affected by the heating associated with photoionization, which would tend to smooth out the gas and so reduce $C$ \citep{pawlik2009}.

\subsection{CMB optical depth}
\label{sec:taucmb}

Once we have the evolution of the mean neutral fraction we can calculate the CMB optical depth directly.
\begin{equation}
\tau_{\rm CMB}=\int_0^{z_{\rm CMB}} \ud z\, \frac{\ud t}{\ud z} [1+f_{\rm He}(z)]Q_{\rm HII}(z) n_H(z) \sigma_T,
\end{equation}
where $\sigma_T$ is the Thompson cross-section and $f_{\rm He}$ is a correction due to the presence of helium and we assume that HeI is ionized at the same time as HI and that HeII reionization occurs at $z=3$.  A more detailed calculation would incorporate the covariance of the measured \tauCMB with other cosmological parameters, but that lies beyond the scope of this paper.

 
 \end{document}